%% file: DarkDiag.tex
\documentclass[10pt,journal]{IEEEtran}
\IEEEoverridecommandlockouts

\usepackage{amsfonts, amsmath,amssymb,amsthm}
\usepackage{algorithm, algorithmic}
\usepackage{array}
\usepackage{cases}
\usepackage{changepage}
\usepackage[compress]{cite}
\usepackage{epstopdf}
\usepackage[english]{babel}
\usepackage{etoolbox}
\usepackage{graphicx}
\usepackage{multirow}
\usepackage{physics} 	
\usepackage{subfigure}
\usepackage{setspace}
\usepackage{url}  
\usepackage[hyphenbreaks]{breakurl} 
\usepackage{lipsum}
\usepackage{nicefrac}	
\usepackage[table,xcdraw]{xcolor}	
\usepackage{textcomp}
\usepackage{makecell}
\usepackage[font=small]{caption} 
\usepackage{mathrsfs}
\usepackage[font=footnotesize]{caption} 

\makeatletter
\patchcmd{\@thm}
  {\trivlist}
  {\list{}{\leftmargin=\thm@leftmargin\rightmargin=\thm@rightmargin}}
  {}{}
\patchcmd{\@endtheorem}
  {\endtrivlist}
  {\endlist}
  {}{}
\newlength{\thm@leftmargin}
\newlength{\thm@rightmargin}

\newcommand{\xnewtheorem}[3]{%
  \newenvironment{#3}
    {\thm@leftmargin=#1\relax\thm@rightmargin=#2\relax\begin{#3INNER}}
    {\end{#3INNER}}%
  \newtheorem{#3INNER}%
}

\makeatother

\newtheoremstyle{indentedupright}{3pt}{3pt}{} {}{\bfseries}{.}{.5em}{} 
\newtheoremstyle{indenteditalic}{3pt}{3pt}{\itshape} {}{\bfseries}{.}{.5em}{} 

\theoremstyle{indenteditalic}
\xnewtheorem{0pt}{0pt}{oldpro}{Conventional Property}
\xnewtheorem{0pt}{0pt}{newpro}{New Property}
\xnewtheorem{0pt}{0pt}{obser}{Observation}
\xnewtheorem{0pt}{0pt}{pro}{Property}
\xnewtheorem{0pt}{0pt}{lem}{Lemma}
\xnewtheorem{0pt}{0pt}{corl}{Corollary}

\newcommand{\romu}[1]{\uppercase\expandafter{\romannumeral #1\relax}} 
\newcommand{\roml}[1]{\lowercase\expandafter{\romannumeral #1\relax}}    

\renewcommand{\baselinestretch}{1}

\begin{document}

\title{Self-Referenced Diagnosis of SPAD Dark-Count-Rate Switching for Quantum Communication Receivers}

\author{Yu Luo, Lina Pu, Yaroslav Koshka, Chun-Hung Liu, Junming Diao%

\thanks{Y. Luo, Y. Koshka and J. Diao are with the Department of Electrical and Computer Engineering, Mississippi State University, Mississippi State, MS 39759 USA (e-mail: yu.luo@ece.msstate.edu).}%
\thanks{Lina Pu is with the Department of Computer Science, University of Alabama, Tuscaloosa, AL 35487 USA (e-mail: lina.pu@ua.edu).}%
\thanks{C.-H. Liu is with the Department of Electrical and Computer Engineering, Texas Tech University, Lubbock, TX 79409 USA (e-mail: Chun-Hung.Liu@ttu.edu)}
\thanks{Corresponding author: Lina Pu.}}


\IEEEtitleabstractindextext{%
\input{Abstract.tex}

}

\maketitle
\IEEEdisplaynontitleabstractindextext
\IEEEpeerreviewmaketitle

\input{Introduction.tex}

\input{Related.tex}

\input{ExpSet.tex}

\input{RtsNoise.tex}

\input{SelfRef.tex}

\input{TmpArt.tex}

\input{EnvArt.tex}

\input{ReadArt.tex}

\input{Conclusion.tex}

\bibliographystyle{IEEEtran}
\bibliography{DarkDiag}

\end{document}

%% file: Abstract.tex
\begin{abstract}
\label{:Abstract}
Single-photon avalanche diodes (SPADs) are essential components of quantum communication receivers. In practical systems, some SPADs exhibit intrinsic step-like changes in dark-count rate (DCR), commonly referred to as random telegraph signal (RTS) noise. Similar RTS-like variations, however, can also arise from stray light, temperature-dependent threshold crossings, cable disturbances, electrical interference, and other readout artifacts. Distinguishing intrinsic RTS from these artifacts is important because they originate from different mechanisms and require different corrective actions. We present a self-referenced diagnostic that primarily uses the low-to-high afterpulse-yield ratio $R$, together with complementary checks based on DCR-state reproducibility, cross-channel behavior, and event timing. Four commercial SPAD modules were monitored in darkness for more than 920\,h. Three exhibited intrinsic DCR switching of 44.59\%, 19.33\%, and 2.77\%, with corresponding $R$ values of $0.9996\pm0.0027$, $1.0050\pm0.0023$, and $1.0027\pm0.0043$, all consistent with unity. Controlled experiments reproduced temperature-related threshold crossings, light leakage, cable-contact artifacts, and artificial electrical injection. These artifacts produced distinct correlation and timing signatures: temperature-related threshold crossings gave intermediate pair-survival responses, cable disturbances and electrical injection generated sub-dead-time events, while light leakage produced simultaneous cross-channel switching and nonreproducible DCR levels. Combining $R$ with orthogonal timing and DCR-based checks provides a practical self-referenced diagnostic framework for distinguishing intrinsic RTS from non-intrinsic artifacts using primarily time-tagged dark-count data.


\end{abstract}

%% file: Introduction.tex
\section{Introduction}
\label{sec:Int}

Single-photon avalanche diodes (SPADs) are key components of quantum communication receivers and photonic quantum technologies, providing single-photon sensitivity and high temporal resolution~\cite{wang2023fast, cusini2022historical2, wojtkiewicz2021review}. When operated above breakdown in Geiger mode, SPADs can also undergo spontaneous avalanches in the absence of incident photons, giving rise to dark counts. The dark-count rate (DCR) is therefore a critical detector parameter, since dark counts contribute to false detections and accidental coincidences and can limit the fidelity and useful operating range of photon-counting systems. Their impact becomes particularly important in quantum links with substantial channel loss, where the received signal approaches the detector background~\cite{liao2017satellite, yin2017satellite}. Reliable characterization of the DCR and its temporal stability is therefore essential for the calibration and maintenance of quantum communication receivers.

The DCR of an SPAD is not necessarily stationary. Some devices exhibit stochastic switching among discrete DCR levels, a phenomenon known as random telegraph signal (RTS) noise~\cite{karami2010rts}. Such behavior is commonly associated with metastable defects or defect complexes whose configurational or charge-state transitions modify local carrier generation and the resulting dark-avalanche rate~\cite{capua2021investigation}. However, a time-to-digital converter (TDC) records electrical pulses that cross its input threshold without identifying their physical origin. Consequently, step-like changes in the recorded DCR can arise not only from intrinsic RTS but also from stray illumination, temperature-dependent threshold crossings, or readout-chain instability. Although intrinsic RTS often exhibits characteristic features, such as abrupt transitions between reproducible DCR levels and approximately exponential dwell-time distributions, similar patterns may occur under certain external disturbances. Thus, DCR behavior alone does not always uniquely identify the switching mechanism. This distinction is important because intrinsic RTS may require detector recalibration or replacement, whereas optical leakage or readout faults require corrective actions elsewhere in the receiver.

In this work, we propose a self-referenced diagnostic framework that uses afterpulse correlations to investigate the origin of DCR switching. Afterpulses are secondary avalanches associated with the delayed release of carriers trapped during preceding avalanches, providing an internal reference for comparing different DCR states. The method primarily uses the low-to-high afterpulse-yield ratio $R$, which captures information about short-delay event correlations beyond the average DCR. Under specific models, $R$ provides distinct predictions for changes in the detected avalanche population, independent counting loss, and uncorrelated count addition. These predictions are combined with complementary checks based on DCR-state reproducibility, cross-channel behavior, and event timing relative to the detector dead-time constraint to distinguish intrinsic RTS from non-intrinsic artifacts.

Conventional diagnostic approaches often involve temperature or bias sweeps, controlled optical excitation, or waveform measurements, which may require additional experimental procedures or instrumentation. In contrast, the proposed method extracts the afterpulse-yield ratio directly from the time-tagged dark-count records already used for DCR monitoring, without requiring dedicated measurements for the initial diagnosis. It therefore provides a complementary, mechanism-sensitive test that can help distinguish different sources of apparent DCR switching. When the afterpulse-yield ratio alone is inconclusive, additional DCR-based, timing, or environmental checks can be used for further diagnosis.

The proposed method was evaluated using four commercial PerkinElmer SPCM-AQR-series SPAD modules in a dark-monitoring campaign comprising more than 920\,h of live acquisition. Three modules exhibited resolved rate states at approximately 177.38\,$\leftrightarrow$\,320.13\,Hz, 296.17\,$\leftrightarrow$\,367.14\,Hz, and 305.7\,$\leftrightarrow$\,314.4\,Hz, corresponding to separations of 44.59\%, 19.33\%, and 2.77\% relative to the high-state rate. Their low-to-high afterpulse-yield ratios were $0.9996\pm0.0027$, $1.0050\pm0.0023$, and $1.0027\pm0.0043$, respectively, all consistent with unity within $2.17\sigma$. Despite the wide range of switching amplitudes, the correlated afterpulse yield is therefore preserved across the resolved DCR states, supporting a detector-local change in the avalanche population.

Controlled perturbations, including stray-light leakage, electrical-pulse injection, temperature-dependent threshold crossings, and cable disturbances, were also examined to establish the signatures and limitations of the diagnostic. These experiments show that different artifacts can produce distinct changes in $R$, cross-channel behavior, or event-timing statistics, while some mechanisms can preserve the afterpulse yield and therefore require additional DCR- or timing-based checks. The resulting framework is intended for detector qualification, dark calibration, and mechanism-specific fault diagnosis in quantum receivers. Because its principal observables are extracted from naturally occurring time-tagged dark-count records, it can be applied without a dedicated optical setup and can help identify the appropriate corrective action for different sources of apparent DCR switching.

The main contributions are summarized as follows:
\vspace{0.1cm}
\begin{adjustwidth}{-0.9cm}{0cm}
\begin{description}
\setlength{\labelsep}{-0.6em}
\itemsep 0.07cm
  \item[(i)] \textbf{A new afterpulse signature of RTS in SPADs is identified.} Three commercial modules exhibit DCR switching amplitudes ranging from approximately 3\% to 45\%, while their low-to-high afterpulse-yield ratios remain consistent with unity. To our knowledge, this approximate preservation of the afterpulse yield across resolved RTS states has not previously been reported.
  \item[(ii)] \textbf{Multiple mechanisms that mimic DCR switching are experimentally reproduced.} Temperature-dependent threshold crossings, stray-light leakage, cable disturbances, and artificial electrical injection produce RTS-like changes with distinct correlation, cross-channel, or event-timing signatures, providing controlled benchmarks for evaluating the diagnostic.
  \item[(iii)] \textbf{A self-referenced diagnostic framework is developed.} The state-resolved afterpulse-yield ratio $R$ provides the primary tests for avalanche-population changes, counting loss, and count addition. DCR-state reproducibility, cross-channel behavior, and event timing relative to the detector dead-time constraint provide complementary discrimination when the correlation metrics alone are insufficient.
  \item[(iv)] \textbf{The method is validated through extended detector monitoring.} A measurement campaign comprising over 920\,h of live acquisition across four commercial SPAD modules demonstrates that intrinsic RTS and non-intrinsic artifacts can be diagnosed primarily from time-tagged dark-count records, supporting detector characterization, calibration, and fault diagnosis in quantum communication receivers.
\end{description}
\end{adjustwidth}
\vspace{0.1cm}

The remainder of this paper is organized as follows. Section~\ref{sec:Related} reviews prior work on SPAD DCR variability, afterpulsing, and RTS noise. Section~\ref{sec:ExpSet} describes the experimental setup, data acquisition, and afterpulse-yield estimation. Section~\ref{sec:RtsNoise} characterizes the observed RTS noise and its state-resolved afterpulsing behavior, while Section~\ref{sec:SelfRef} develops the self-referenced diagnostic. Section~\ref{sec:TmpArt} examines temperature-related RTS-like artifacts. Section~\ref{sec:EnvArt} investigates environment-induced artifacts, including light leakage and cable disturbances. Section~\ref{sec:ReadArt} discusses readout-dependent artifacts, including counting addition and counting loss. Section~\ref{sec:Con} concludes the paper.

%% file: Related.tex
\section{Related Work}
\label{sec:Related}

Discrete switching of SPAD dark-count rates has been reported in both unirradiated and irradiated devices. Early studies documented two-level and multilevel random telegraph signal (RTS) behavior and characterized its dependence on temperature and excess bias~\cite{karami2010rts}. Studies of irradiated SPADs linked RTS to radiation-induced generation centers and metastable defect configurations~\cite{di2018random,capua2021investigation}, while measurements of CMOS SPADs examined its voltage and temperature dependence and its association with depletion-region defects~\cite{jiang2021random}. Together, these findings support defect-mediated explanations for discrete DCR states. They do not, however, establish that every transition in a recorded count stream originates within the SPAD.

The rate recorded by a time-to-digital converter (TDC) depends on both avalanche generation and the subsequent quenching, recovery, pulse transmission, and discrimination processes. Changes in excess bias or breakdown voltage may alter avalanche generation, whereas changes in recovery, discriminator threshold, or readout conditions may alter which output pulses are registered~\cite{cova1996avalanche,mahmoudi2021noise}. Stray light can likewise change the recorded rate without changing the intrinsic dark-count rate. A step-like transition in the TDC record therefore does not, by itself, identify its cause. Conventional diagnosis often requires supplementary measurements or controlled tests, such as bias and temperature sweeps, optical excitation, waveform acquisition, or environmental monitoring~\cite{cusini2022historical1}.

Afterpulsing has been extensively characterized as an SPAD noise mechanism. Established methods use photon-counting statistics~\cite{zhang2008photon}, deviations from Poisson statistics~\cite{tzou2015method}, interarrival-time distributions~\cite{hofbauer2018temperature}, and correlations in time-tagged events~\cite{wayne2017simple}. Further studies have examined biases in afterpulse measurements~\cite{dejen2024refined}, the effects of excess bias and gating~\cite{mahmoudi2021noise}, behavior following irradiation~\cite{wu2022radiation}, and afterpulsing in Ge-on-Si and Ge--Si SPADs~\cite{yi2023afterpulsing,na2024room}. These studies characterize afterpulsing across devices and operating conditions, but generally do not compare its correlated yield between naturally occurring high- and low-DCR states of the same detector. To our knowledge, this state-conditioned yield has not previously been used to diagnose the origin of a count-rate transition.

Here, we use afterpulse correlations as a self-referenced diagnostic. Comparing the correlated afterpulse yield across rate states tests whether the statistical relationship between avalanches and subsequent afterpulses is preserved; a normalized pair-survival statistic provides a complementary test for correlated counting loss. Together with cross-channel and waveform controls, these measurements help distinguish detector-local switching from the external perturbations and readout artifacts examined in this work, while relying primarily on time-tagged dark-count data.

%% file: ExpSet.tex
\section{Experimental Setup and Data Acquisition}
\label{sec:ExpSet}
This section describes the experimental setup, beginning with the system hardware for self-referenced diagnosis, followed by the data acquisition and afterpulse measurement procedures.

\subsection{System Hardware and Data Acquisition}
\label{subsec:HarSet}
\begin{figure}[htb]
\centerline{\includegraphics[width=\linewidth]{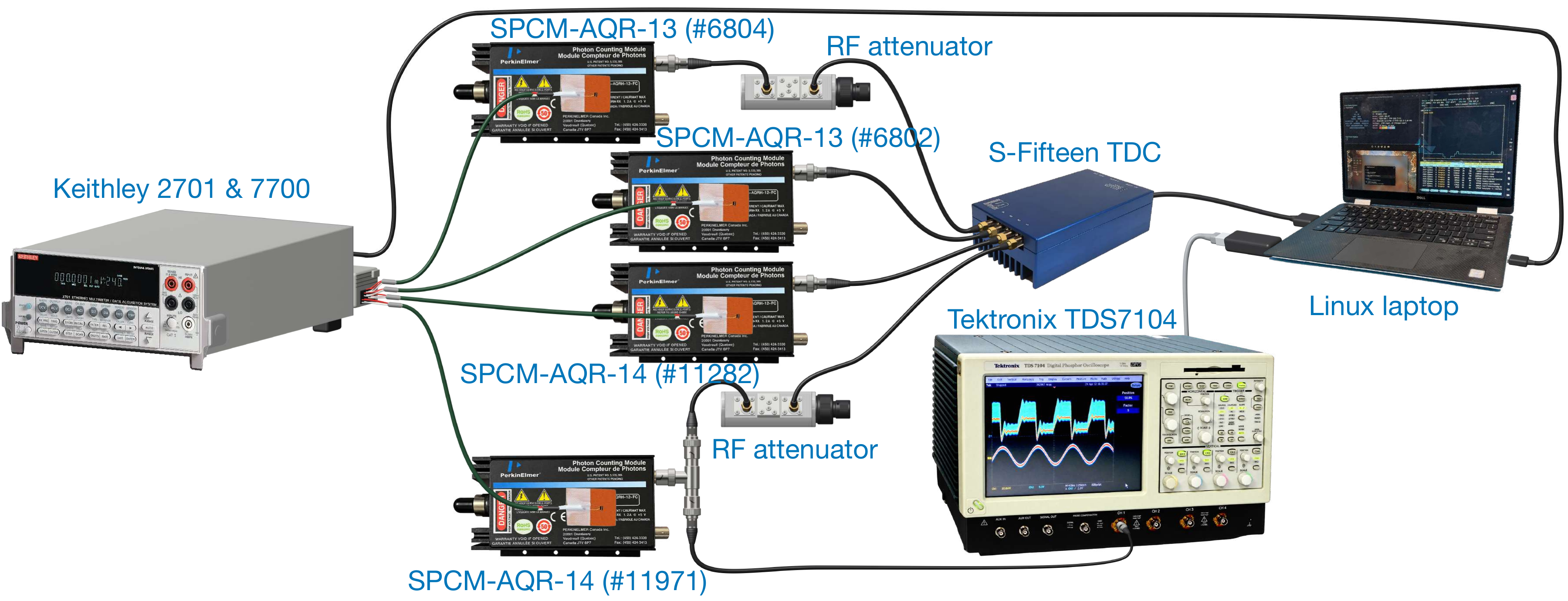}}
  \caption{Measurement system for self-referenced diagnosis, where RTD sensors are attached to the SPAD module housings using orange tape.}\label{fig:ExpSet}
\end{figure}

The experimental setup for the self-referenced diagnosis is shown in Fig.~\!\ref{fig:ExpSet}. Four PerkinElmer SPCM-AQR SPAD modules are connected via coaxial cables to a four-channel S-Fifteen TDC with a timing resolution of 2\,ns. The optical entrance of each module is individually sealed with opaque metal foil. The outputs of SPCM-AQR-13 \#6802 and SPCM-AQR-14 \#11282 are connected directly to the TDC. The outputs of the remaining modules, SPCM-AQR-13 \#6804 and SPCM-AQR-14 \#11971, each pass through a JFW 50DR-096 variable step attenuator (0--30\,dB in 1\,dB steps, DC--3\,GHz). The output of SPCM-AQR-14 \#11971 is also connected to a Tektronix TDS7104 oscilloscope via a BNC T-adapter for waveform measurements when needed. By reducing pulse amplitudes relative to the TDC input threshold, the attenuators produce controlled, amplitude-selective counting losses. Measurements of the corresponding pulse-height distributions help assess how these losses affect afterpulse measurements and mimic RTS-like count-rate changes, as detailed in Section~\!\ref{sec:TmpArt}. Unless otherwise noted, all intrinsic-RTS measurements were acquired with the TDC input threshold set to 1.599\,V TTL level on all four channels and with every step attenuator at 0\,dB.

Five Omega SA1-RTD-4W surface-mount RTD sensors monitor the temperatures of the four SPAD module housings and the ambient room. These Class A Pt100 sensors have a specified tolerance of approximately $\pm0.20\,^{\circ}\mathrm{C}$ at $25\,^{\circ}\mathrm{C}$ and less than $0.20\,^{\circ}\mathrm{C}$ drift per year. They are read out in a four-wire configuration by a Keithley 2701 multimeter/data-acquisition system equipped with a 7700 multiplexer module. Each sensor is sampled once per minute. The TDC, oscilloscope, and Keithley 2701 are connected to a laptop running Linux for data collection and processing.

The TDC records each registered SPAD output pulse in a 32-bit word comprising a 27-bit time counter with 2\,ns resolution, a marker bit, and a 4-bit channel mask. Pulses on different channels within the same time bin share one word. Records are grouped into 60\,s chunks, each labeled with the host's NTP-synchronized Unix timestamp. Counter rollovers are corrected using the combined four-channel record. Inter-arrival times and correlations are calculated solely from TDC timestamps within each chunk, while host timestamps align the chunks with temperature measurements. No timing statistics span the acquisition gaps between chunks.

\subsection{Afterpulse Identification}
\label{subsec:AftIden}
\begin{figure}[htb]
\centerline{\includegraphics[width=6.5cm]{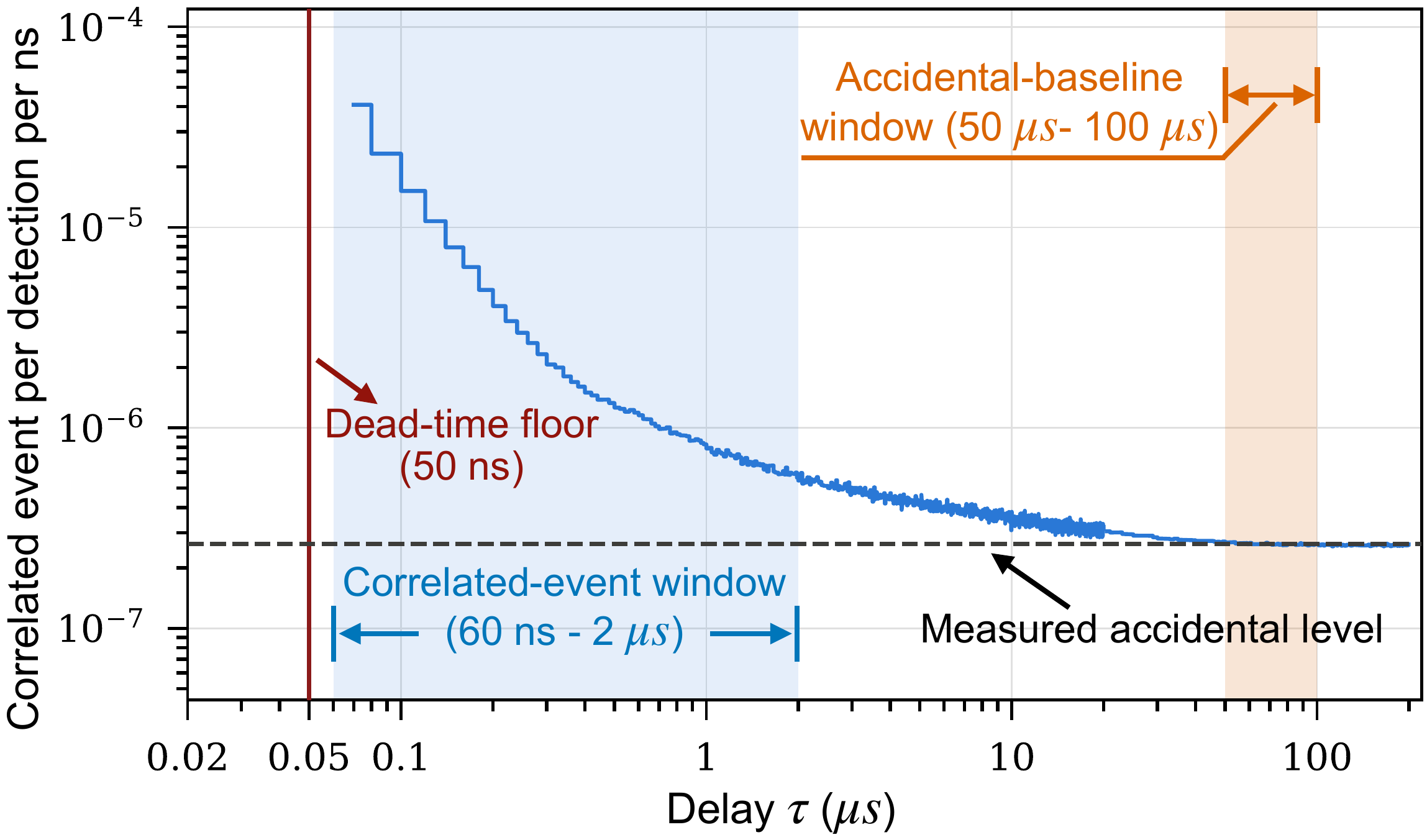}}
\caption{Identification of the afterpulse population in SPCM-AQR-14 \#11971.}
\label{fig:AfterMea}
\end{figure}

In the experiment, afterpulsing is identified statistically as an excess of short-delay event pairs above the uncorrelated background. Fig.~\!\ref{fig:AfterMea} illustrates the identification procedure using the SPCM-AQR-14 module \#11971 as an example. For each detector, the dead-time floor is defined as the shortest inter-event delay observed in the recorded timestamp data. Below this delay, no events are recorded because the module remains temporarily insensitive while recovering from the preceding avalanche. For \#11971, the measured dead-time floor is approximately 50\,ns. The blue curve is the measured forward all-pairs delay density, i.e., correlogram. Each event pair consists of a recorded detection and a subsequent detection separated by a delay $\tau$.

The blue shaded region in Fig.~\!\ref{fig:AfterMea} is the short-delay integration window, spanning 60\,ns to 2.00\,$\mu$s, and the number of event pairs within this window is denoted by $N_s$. The lower bound is set at 60\,ns rather than at the approximately 50\,ns dead-time floor because the histogram uses 20\,ns bins. Thus, 60\,ns is the lower edge of the first bin lying entirely above the measured dead time, avoiding a bin that is only partially observable because of detector dead time. The upper bound of 2.00\,$\mu$s captures the dominant short-delay correlated population, which decays on approximately the microsecond timescale, while limiting the contribution from unrelated pairs at longer delays.

The orange accidental-baseline window, spanning 50--100\,$\mu$s, is chosen where the correlogram has become flat and the short-delay correlated excess has decayed away. The mean pair density in this region therefore provides the measured accidental level, representing the density of uncorrelated event pairs. This measured density is scaled by the width of the short-delay integration window and by the number of recorded detections to obtain $N_b$, the expected number of accidental event pairs within that window. The absolute afterpulse yield, denoted by $n_{\mathrm{ap}}$, is then
\begin{equation}
\label{eq:nshu}
n_{\mathrm{ap}}=\frac{N_s-N_b}{N_d},
\end{equation}
where $N_d$ is the total number of recorded detections.

%% file: RtsNoise.tex
\section{RTS Noise and Afterpulsing in SPADs}
\label{sec:RtsNoise}
This section examines the characteristics of RTS noise and afterpulsing in SPADs. We first characterize the RTS behavior observed in three SPAD modules and then investigate the afterpulsing characteristics extracted from the experimental data.

\subsection {Characteristics of RTS Noise}
\label{subsec:CharRts}
\begin{figure}[htb]
\centerline{\includegraphics[width=\linewidth]{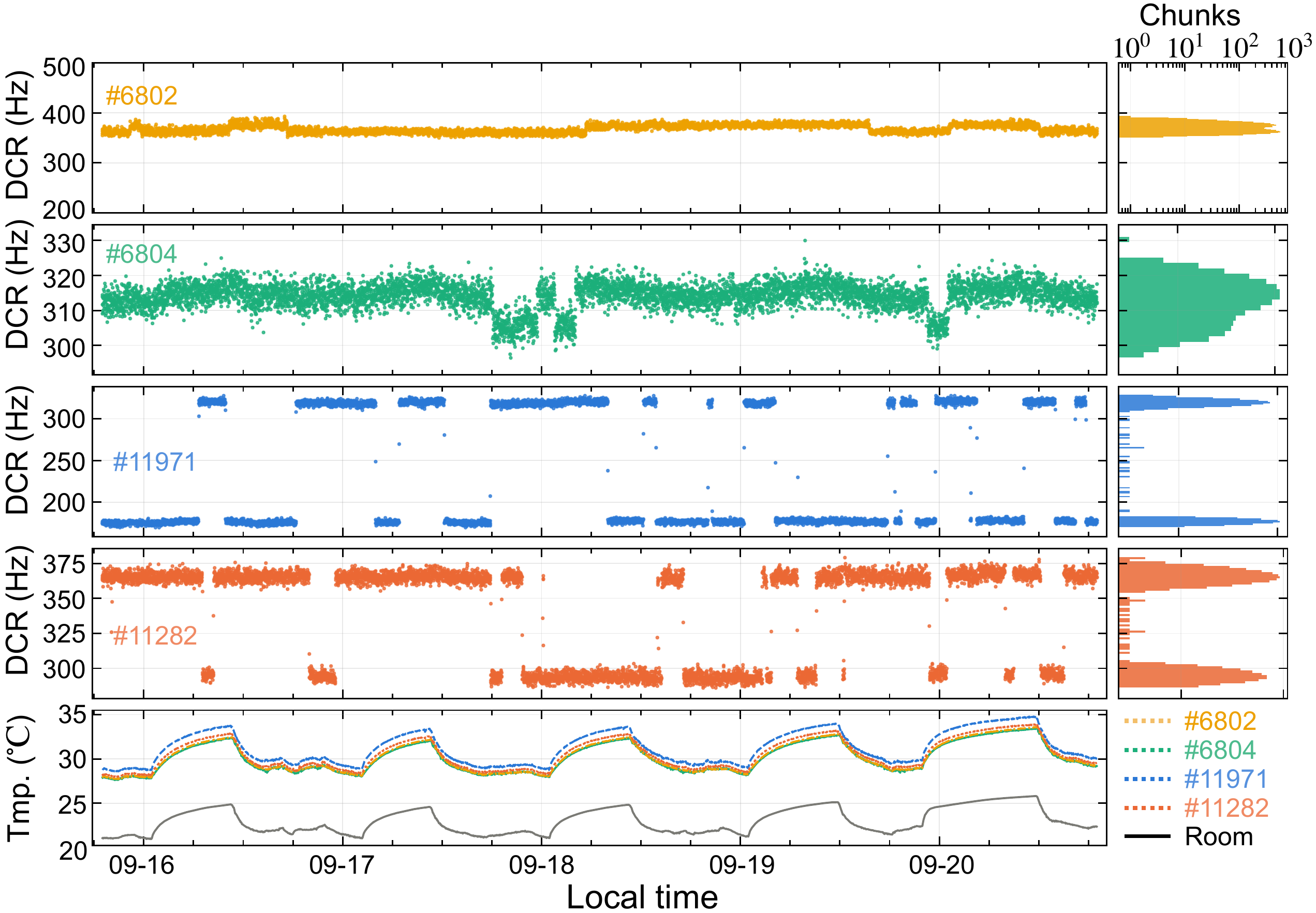}}
  \caption{Segment of the dark-count record, with local time reported in U.S. Central Time (2026).}\label{fig:RtsMea}
\end{figure}

Fig.~\!\ref{fig:RtsMea} shows a four-day segment of the module-housing temperature and DCR for the four SPAD modules, measured by the RTD sensors and TDC, respectively. The DCR records exhibit slow baseline fluctuations in all four modules, while three modules additionally show resolved discrete rate states. Modules SPCM-AQR-14 \#11971 and \#11282 undergo abrupt switching between well-separated levels with irregular dwell times, producing clear bimodal rate histograms. Their low- and high-rate states are approximately 177.38\,$\leftrightarrow$\,320.13\,Hz and 296.17\,$\leftrightarrow$\,367.14\,Hz, corresponding to separations of 44.59\% and 19.33\%, respectively, relative to the high-state rate. Module SPCM-AQR-13 \#6804 exhibits much smaller intermittent downward steps, with resolved levels near 305.7\,$\leftrightarrow$\,314.4\,Hz; these steps are distinguishable in the time record but not in the marginal rate histogram. In contrast, \#6802 shows no resolved discrete state. The three resolved state pairs therefore span a wide range of relative separations, from approximately 3\% to 45\%.

By analyzing the experimental data, we find that the dwell-time statistics are consistent with stochastic, approximately memoryless switching. For \#11971, the mean dwell times are 2.66\,h in the low state and 4.65\,h in the high state, while for \#11282 they are 2.74\,h and 7.22\,h, respectively. For these four sets of dwell times, the coefficients of variation range from 0.95 to 1.06, and none of the corresponding distributions shows a significant deviation from an exponential distribution. Successive dwell times are also only weakly correlated, with no statistically significant dependence, indicating that the duration of one dwell provides little information about the next.

Although DCR of an SPAD can depend on temperature~\cite{hofbauer2018temperature}, the abrupt rate transitions neither consistently coincide with the smooth temperature cycles nor occur simultaneously across modules, indicating that gradual common-mode thermal variation does not account for the observed switching. The substantial differences in absolute DCR among modules of the same commercial series further emphasize that random RTS noise is identified from discrete changes within an individual device rather than from its absolute DCR relative to another module. 

To further resolve the switching timescale, each clean transition in \#11971 was fitted using a Poisson rate model in which the two rate levels were connected by a linear ramp of duration $\tau$. No finite transition ramp was resolved. For the 125 low-to-high and 125 high-to-low transitions, the pooled 95\% upper limits on $\tau$ are 42\,$\mu$s and 37\,$\mu$s, respectively. Combining all 250 transitions gives a tighter pooled 95\% upper limit of $\tau\!<\!21$\,$\mu$s. The same analysis was applied to \#11282, yielding 200 clean transitions, including 99 low-to-high and 101 high-to-low transitions. The corresponding pooled 95\% upper limits are 0.48\,ms and 0.38\,ms, respectively, while combining all transitions gives $\tau\!<\!0.17$\,ms. The less stringent limit for \#11282 reflects its smaller DCR switching amplitude and the fewer available transitions.

The above limits are determined primarily by the stochastic dark-count statistics rather than the 2\,ns timing resolution of the TDC. At a representative count rate of 300\,Hz, the expected number of detections within 20\,$\mu$s is only 0.006, making short transition ramps difficult to resolve. Therefore, the results indicate that the observed switching is consistent with abrupt transitions, but do not imply that the physical transition duration is zero or exclude faster switching dynamics.

\subsection{Afterpulsing Across Rate States}
\label{subsec:CharAfter}

The absolute afterpulse yield, $n_{\mathrm{ap}}$, varies among modules and is not constant over time. Its magnitude differs substantially across the measured devices, ranging from approximately $0.18\%$ for \#6804 to $0.33\%$ and $0.45\%$ for \#11971 and \#11282, respectively. In addition, $n_{\mathrm{ap}}$ exhibits a weak temperature dependence: within-day fits give relative temperature coefficients ranging from  $0.07\pm0.06$ to $0.25\pm0.05\%\,\mathrm{K}^{-1}$ in the high-DCR state and from $0.02\pm0.09$ to $0.19\pm0.09\%\,\mathrm{K}^{-1}$ in the low-DCR-state.

A slower long-term drift in $n_{\mathrm{ap}}$ is also observed. Over 44 daily measurements, $n_{\mathrm{ap}}$ decreases by $2.0\%$--$7.5\%$ across the four modules. The largest change occurs in \#11971, where the yield decreases from 0.3588\% to 0.3317\%, whereas the smallest occurs in \#6804, from 0.1840\% to 0.1802\%. Thus, the absolute value of $n_{\mathrm{ap}}$ is neither common to nominally similar modules nor sufficiently stable in time to serve as an accurate reference.

In contrast, $n_{\mathrm{ap}}$ is preserved between the resolved low- and high-DCR states of each module. We quantify this state-to-state preservation using the low-to-high afterpulse-yield ratio,
\begin{equation}
\label{ieq:9sg}
R=\frac{n_{\mathrm{ap}}^{\mathrm{L}}} {n_{\mathrm{ap}}^{\mathrm{H}}}.
\end{equation}

Throughout the entire experimental period (44 days), $R=0.9996\pm0.0027$, $1.0050\pm0.0023$, and $1.0027\pm0.0043$ for \#11971, \#11282, and \#6804, respectively. All three ratios are statistically consistent with unity,  despite corresponding DCR changes of $44.59\%$, $19.33\%$, and $2.77\%$. Moreover, $R$ remains stable over the long-term measurement, with no statistically significant slow drift or temperature-dependent variation within the measurement uncertainty, even though the absolute afterpulse yield varies with both time and temperature.

These results show that although the absolute afterpulse yield varies with time and temperature, the afterpulse yield per recorded count is preserved across transitions between resolved DCR states, while common-mode variations in the absolute yield largely cancel in the low-to-high ratio. Therefore, the state-to-state invariance and long-term stability of $R$, rather than the absolute magnitude of $n_{\mathrm{ap}}$, provide the self-reference used in Section~\ref{sec:SelfRef} to diagnose the origin of DCR variations in quantum communication receivers.

%% file: SelfRef.tex
\section{Self-Referenced Diagnostic Method}
\label{sec:SelfRef}

This section develops criteria for diagnosing DCR switching. The afterpulse-yield ratio tests avalanche-population changes against independent count loss and uncorrelated count addition, while a normalized pair-survival statistic assesses correlated loss. We then combine these measures with auxiliary checks and discuss their limitations.

\subsection{Rate- and Yield-Ratio Tests}

Let $r_d^{\mathrm{L}}$ and $r_d^{\mathrm{H}}$ denote the low- and high-DCR recorded by the TDC, respectively. Their ratio is denoted by $p$, which is 
\begin{equation}
\label{eq:swdv}
	p=\frac{r_d^{\mathrm{L}}}{r_d^{\mathrm{H}}},
	\qquad 0<p<1,
\end{equation}
so the fractional rate step relative to the high state is $1-p$. Let $N_d^i$ and $N_c^i$ denote the numbers of recorded detections and background-subtracted correlated pairs in state $i\in\{\mathrm{L},\mathrm{H}\}$. According to (\ref{eq:nshu}) and (\ref{ieq:9sg}), it can be obtained that 
\begin{equation}
\label{eq:ccr5}
	R=\frac{n_{\mathrm{ap}}^{\mathrm{L}}}{n_{\mathrm{ap}}^{\mathrm{H}}}
	=\frac{N_c^{\mathrm{L}}/N_d^{\mathrm{L}}}{N_c^{\mathrm{H}}/N_d^{\mathrm{H}}}.
\end{equation}

$R$ compares correlated events per recorded detection in the two states, rather than the total number of afterpulses generated by an SPAD. A value $R\simeq1$ means that the two states have the same afterpulse signature per detection, even though their total numbers of detections and correlated pairs may differ. Values below or above unity indicate that the low-rate state contains, respectively, fewer or more correlated events per detection than the high-rate state. Because both yields are measured from the same detector with the same timing windows, their ratio also suppresses slow changes common to both states. In the scaling arguments below, counts are normalized to the same live time.

The expected $R$ depends on how the count rate changes. For a change in the avalanche population, each additional primary avalanche has the same probability of producing a recorded afterpulse. The detection and correlated-pair populations therefore scale together: $N_d^{\mathrm{L}}\simeq pN_d^{\mathrm{H}}$ and $N_c^{\mathrm{L}}\simeq pN_c^{\mathrm{H}}$. According to (\ref{eq:ccr5}), their quotient is preserved, giving $R\simeq1$.

For independent counting loss, the low-rate record is obtained by retaining each output pulse with probability $p$. The total detection count scales as $N_d^{\mathrm{L}}\simeq pN_d^{\mathrm{H}}$, whereas a correlated pair is recorded only if both its primary pulse and afterpulse survive, giving $N_c^{\mathrm{L}}\simeq p^2N_c^{\mathrm{H}}$. Therefore, $n_{\mathrm{ap}}^{\mathrm{L}}\simeq p n_{\mathrm{ap}}^{\mathrm{H}}$ and $R\simeq p$.

For uncorrelated count addition, the high-rate state is modeled as the low-rate record plus events that contribute no short-delay correlated excess. The correlated-pair count remains $N_c^{\mathrm{H}}\simeq N_c^{\mathrm{L}}$, while the total count increases to $N_d^{\mathrm{H}}\simeq N_d^{\mathrm{L}}/p$. Hence, $n_{\mathrm{ap}}^{\mathrm{H}}\simeq p n_{\mathrm{ap}}^{\mathrm{L}}$ and $R\simeq1/p$. The three predictions are therefore
\begin{equation}
\label{eq:o8as}
	R \simeq
	\begin{cases}
		1,   & \text{avalanche-population change},\\
		p,   & \text{independent counting loss},\\
		1/p, & \text{uncorrelated count addition}.
	\end{cases}
\end{equation}
A measured $R\simeq1$, together with statistically significant rejection of $R=p$ and $R=1/p$, therefore excludes these independent-loss and uncorrelated-addition models. It supports a change in the detected avalanche population, but is not sufficient by itself: correlated pair loss and added photon-induced avalanches can also preserve the afterpulse yield and are addressed by the tests below.

\subsection{Normalized Pair-Survival Statistic}
\label{sec:NorPair}

Some counting artifacts may retain or reject a primary pulse and its associated afterpulse together. To quantify this possibility, let $T_p^i$ be the live time in state $i\in\{\mathrm{L},\mathrm{H}\}$ and define the correlated-pair rate as $c_p^i=N_c^i/T_p^i$. The afterpulse yield in (\ref{eq:nshu}) can then be written as
\begin{equation}
\label{eq:werv}
n_{\mathrm{ap}}^i=\frac{N_c^i}{N_d^i}=\frac{N_c^i/T_p^i}{N_d^i/T_p^i}=\frac{c_p^i}{r_d^i}.
\end{equation}

Define the pair-rate ratio as $q=c_p^{\mathrm{L}}/c_p^{\mathrm{H}}$. From \eqref{eq:ccr5} and \eqref{eq:werv},
\begin{equation}
\label{eq:iolm}
	R=\frac{c_p^{\mathrm{L}}/c_p^{\mathrm{H}}}{r_d^{\mathrm{L}}/r_d^{\mathrm{H}}}=\frac{q}{p},
\end{equation}
and therefore $q=pR$. This identity follows directly from the definitions of $R$ and $p$ and remains valid when the two states have different live times.

Now consider a pure counting-loss model in which the low-state readout filters an otherwise unchanged high-state pulse stream. Let $X$ and $Y$ be Bernoulli indicators describing whether a primary pulse and its associated afterpulse, respectively, survive the filtering ($1$: survive; $0$: lost). First consider the equal-marginal case, $P(X=1)=P(Y=1)=p$. Under this model, the pair-rate ratio is the joint survival probability, $q=P(X=1,Y=1)$. The Pearson correlation between the two survival indicators is therefore
\begin{equation}
\label{eq:zapd}
	\rho_{XY}
	=\frac{\operatorname{E}[XY]-\operatorname{E}[X]\operatorname{E}[Y]}{\sqrt{\operatorname{Var}(X)\operatorname{Var}(Y)}}=\frac{q-p^2}{p(1-p)}.
\end{equation}
Substituting  \eqref{eq:iolm} into \eqref{eq:zapd} gives
\begin{equation}
\label{eq:nbv3}
	\rho_{XY}=\frac{R-p}{1-p}.
	\end{equation}
Thus, under the equal-marginal pure-loss model, the survival correlation can be obtained directly from the measured ratios $R$ and $p$. For clarity, we denote the correlation inferred under this model by $\rho_s$, such that $\rho_s=(R-p)/(1-p)$. This quantity is referred to as the normalized pair-survival statistic.

Independent survival between primary pulse and its associated afterpulse gives $q=p^2$, and hence $R=p$ and $\rho_s=0$. Positive but incomplete correlation gives $p<R<1$ and $0<\rho_s<1$, whereas identical survival outcomes give $q=p$, $R=1$, and $\rho_s=1$. Equal marginal survival probabilities alone do not imply identical survival outcomes. Similarly, $R<p$ gives $\rho_s<0$, corresponding to anticorrelated survival within this model.

The equal-marginal assumption need not hold if primary pulses and afterpulses respond differently to the counting loss. More generally, let $p_x=P(X=1)$ and $p_y=P(Y=1)$. The measured rate ratio $p$ is then the survival fraction of the overall recorded pulse population and need not equal either $p_x$ or $p_y$. Under the same pure-filtering model, $q=pR$ remains the joint survival probability, and the physical Pearson correlation becomes
\begin{equation}
\label{eq:avve}
	\rho_{XY}=\frac{pR-p_xp_y}{\sqrt{p_x(1-p_x)p_y(1-p_y)}}.
\end{equation}
For independent survival, $q=p_xp_y$, giving $R=p_xp_y/p$, which reduces to $R\simeq p_y$ only when $p\simeq p_x$. Furthermore, perfect positive correlation, $\rho_{XY}=1$, requires identical Bernoulli survival outcomes and therefore $p_x=p_y$.

Consequently, $\rho_s$ in \eqref{eq:nbv3} is equal to the physical survival correlation $\rho_{XY}$ only under the equal-marginal pure-loss model. If the marginal survival probabilities differ, recovering $\rho_{XY}$ requires separate estimates of $p_x$ and $p_y$, whereas $\rho_s$ remains directly calculable from the measured quantities $R$ and $p$. More generally, outside the pure-loss model, $\rho_s$ should be interpreted only as a normalized pair-survival diagnostic and need not be constrained to the Pearson-correlation interval $[-1,1]$. In particular, for a genuine change in avalanche generation, no underlying high-state pulse stream is being filtered. In that case, $\rho_s$ remains a diagnostic statistic for comparison with counting-loss mechanisms, but it does not represent a physical correlation between pulse-survival outcomes.

\subsection{Decision Criteria and Limitations}
\begin{figure}[htb]
\centerline{\includegraphics[width=\linewidth]{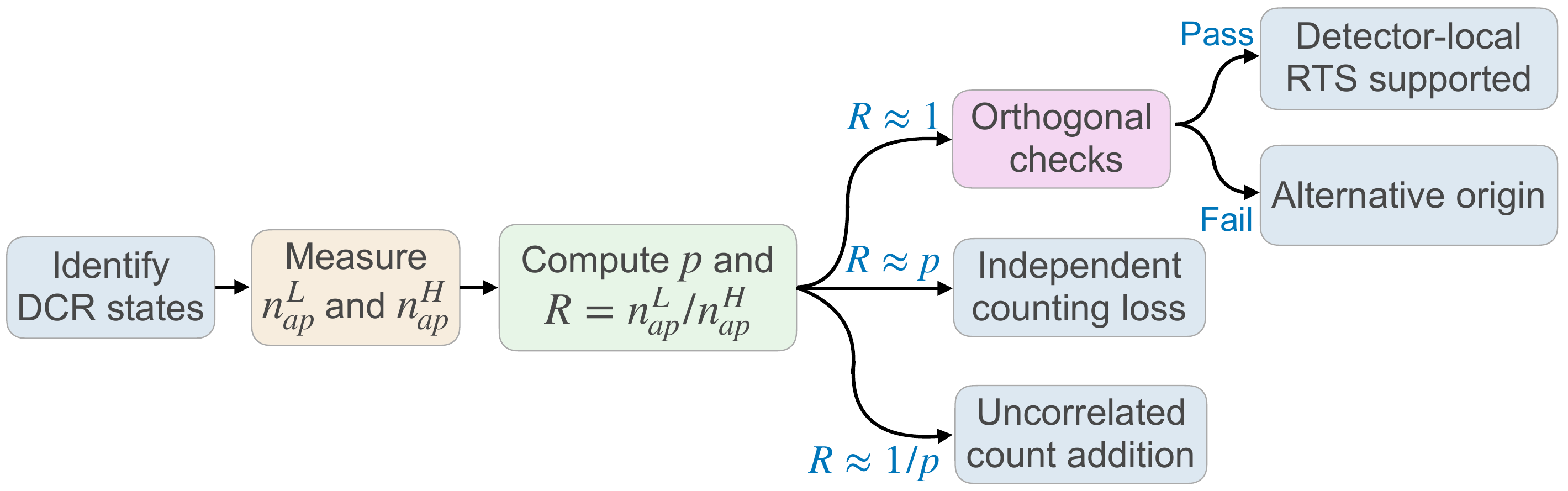}}
  \caption{Flow diagram of main diagnostic procedure.}\label{fig:DiagFlow}
\end{figure}

Based on the relationships among $p$, $R$, and $\rho_s$ established above, together with the orthogonal experimental checks, the main diagnostic procedure for DCR switching in quantum communication receivers can be summarized in Fig.~\!\ref{fig:DiagFlow}. In the following, we describe each diagnostic step in detail and then discuss the limitations of the proposed method.

First, candidate DCR states are identified from recurrent, abrupt transitions between locally stable count rates. Temperature records and simultaneous measurements from other channels are then examined to determine whether the transitions are associated with thermal drift or common-mode disturbances. Second, the rate ratio $p$ and afterpulse-yield ratio $R$ are calculated using identical correlation windows in both states. The measured $R$ is compared with the predictions for an avalanche-population change ($R=1$), independent counting loss ($R=p$), and uncorrelated count addition ($R=1/p$). Complete dwells or measurement blocks are resampled when estimating uncertainties so that temporal variability is preserved. Third, when correlated counting loss remains a plausible explanation, the normalized pair-survival statistic $\rho_s$ is calculated and compared with controlled threshold-loss measurements.

Within the equal-marginal pure-loss model, $\rho_s$ describes how strongly the survival of a primary pulse is associated with the survival of its afterpulse. Independent survival gives $\rho_s=0$, partially correlated survival gives $0<\rho_s<1$, and identical survival outcomes give $\rho_s=1$. Thus, a value below unity indicates that the two pulses are not always retained or rejected together, whereas $\rho_s\simeq1$ is consistent with nearly identical survival. Outside the pure-loss model, $\rho_s$ remains a useful normalized pair-survival statistic but should not be interpreted as a physical Pearson correlation.

A measured $R\simeq1$ that is inconsistent with $R=p$ and $R=1/p$ excludes the independent-loss and uncorrelated-addition models. However, $R\simeq1$ alone is not sufficient to establish detector-local avalanche switching. Perfectly correlated counting loss can also preserve the afterpulse yield and produce $\rho_s\simeq1$, while stray light can generate additional avalanches together with their associated afterpulses and therefore also yield $R\simeq1$.

These remaining ambiguities are addressed using the normalized pair-survival statistic together with orthogonal experimental observations. Evidence for detector-local avalanche switching is obtained when the following observations are jointly satisfied: (i) $R$ is consistent with unity but inconsistent with $p$ and $1/p$; (ii) the DCR transitions are abrupt, occur at random times, and are confined to a single detector channel; (iii) The DCR remains stable within each state, while the dwell times in all states are consistent with exponential distributions; and (iv) the transitions show no correlation with external factors, such as temperature, stray light, or electromagnetic interference, and no corresponding change is observed in the pulse-height distribution. Taken together, these observations reject the tested independent-loss, uncorrelated-addition, common-mode, thermal, and pulse-amplitude mechanisms and support detector-local avalanche switching.

It is worth noting that the discrimination capability decreases as $p\rightarrow1$, because the predictions $R\!=\!1$, $R\!=\!p$, and $R\!=\!1/p$ converge. Statistical consistency should therefore be assessed using confidence intervals or hypothesis tests rather than fixed thresholds, with the effectiveness of the diagnostic depending on the DCR separation and measurement uncertainty.

%% file: TmpArt.tex
\section{Temperature-Related Artifacts}
\label{sec:TmpArt}

A TDC records an event only when the SPAD output crosses its input threshold. The recorded count rate can therefore change even when the underlying avalanche rate is unchanged, if the pulse-amplitude distribution moves relative to the threshold. This effect becomes important when the discriminator is placed near the top of the output pulse, where a small amplitude shift can change whether a pulse crosses the threshold zero, one, or even multiple times. Temperature-dependent pulse amplitudes can consequently produce RTS-like changes in the recorded DCR.

To reproduce RTS-like threshold-crossing artifacts, a step attenuator was inserted between each SPAD output and the TDC, with the attenuation adjusted individually for each module to bring its output amplitude closer to the nominal SPCM-AQR level of 2.5\,V into a 50\,$\Omega$ load specified by the manufacturer~\cite{Excelitas2020spcm}. For \#11971, whose unattenuated output amplitude is approximately 4.8\,V, an attenuation of 6\,dB was applied. This module-specific adjustment reduces device-to-device amplitude variation and provides more consistent electrical conditions for reproducing RTS-like threshold-crossing artifacts.

\begin{figure}[htb]
\centerline{\includegraphics[width=\linewidth]{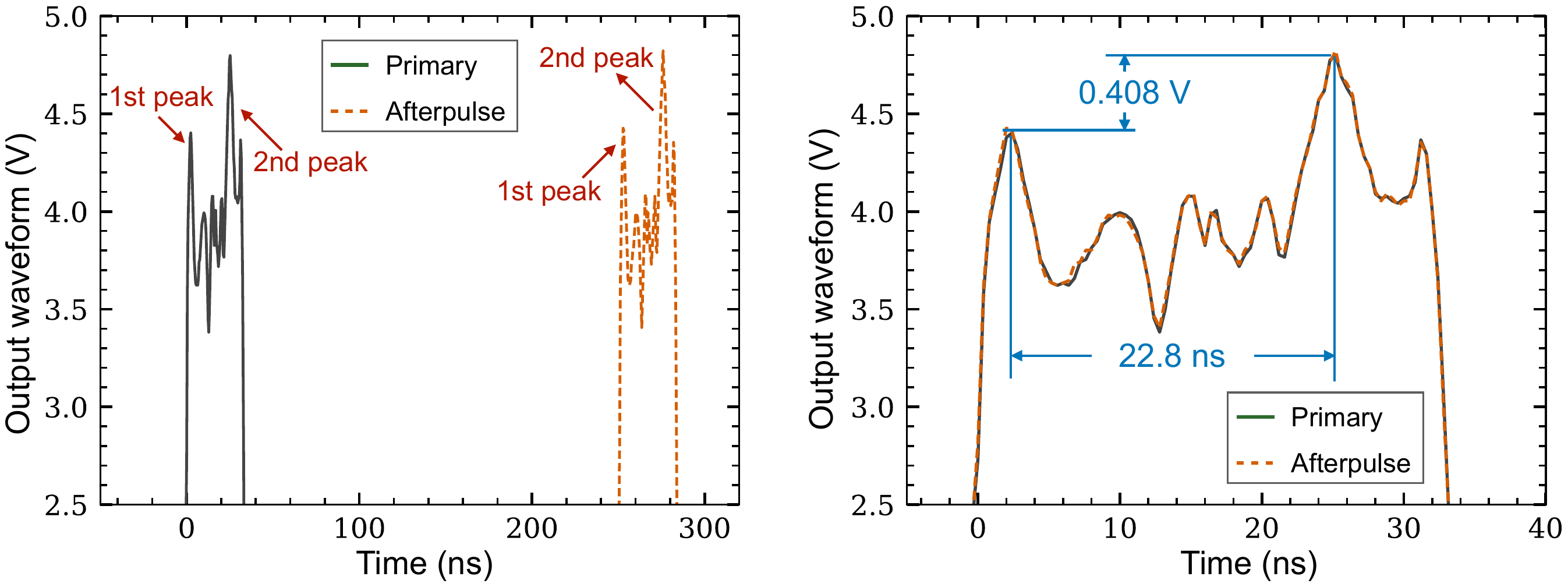}}
\caption{Output waveforms of a primary pulse and its afterpulse from \#11971. (Left) Original waveforms. (Right) Afterpulse waveform shifted in time to align with the primary.}
\label{fig:PriAfrWav}
\end{figure}

We first examine the waveforms of a primary pulse and its associated afterpulse, as shown in Fig.~\!\ref{fig:PriAfrWav}. Both pulses are approximately 33\,ns wide and exhibit nearly identical shapes after temporal alignment. Notably, the pulse top is not flat but contains two local maxima separated by a central dip. The two maxima differ in amplitude by approximately 0.408\,V and are separated by 22.8\,ns, much longer than the 2\,ns timing resolution of the S-Fifteen TDC. Consequently, a threshold well below the pulse top registers each pulse reliably, whereas a threshold placed near the top becomes sensitive to pulse-amplitude variations and may produce either missed counts or two resolvable threshold crossings from a single pulse. These effects are examined in the following analysis.

\begin{figure}[htb]
\centerline{\includegraphics[width=\linewidth]{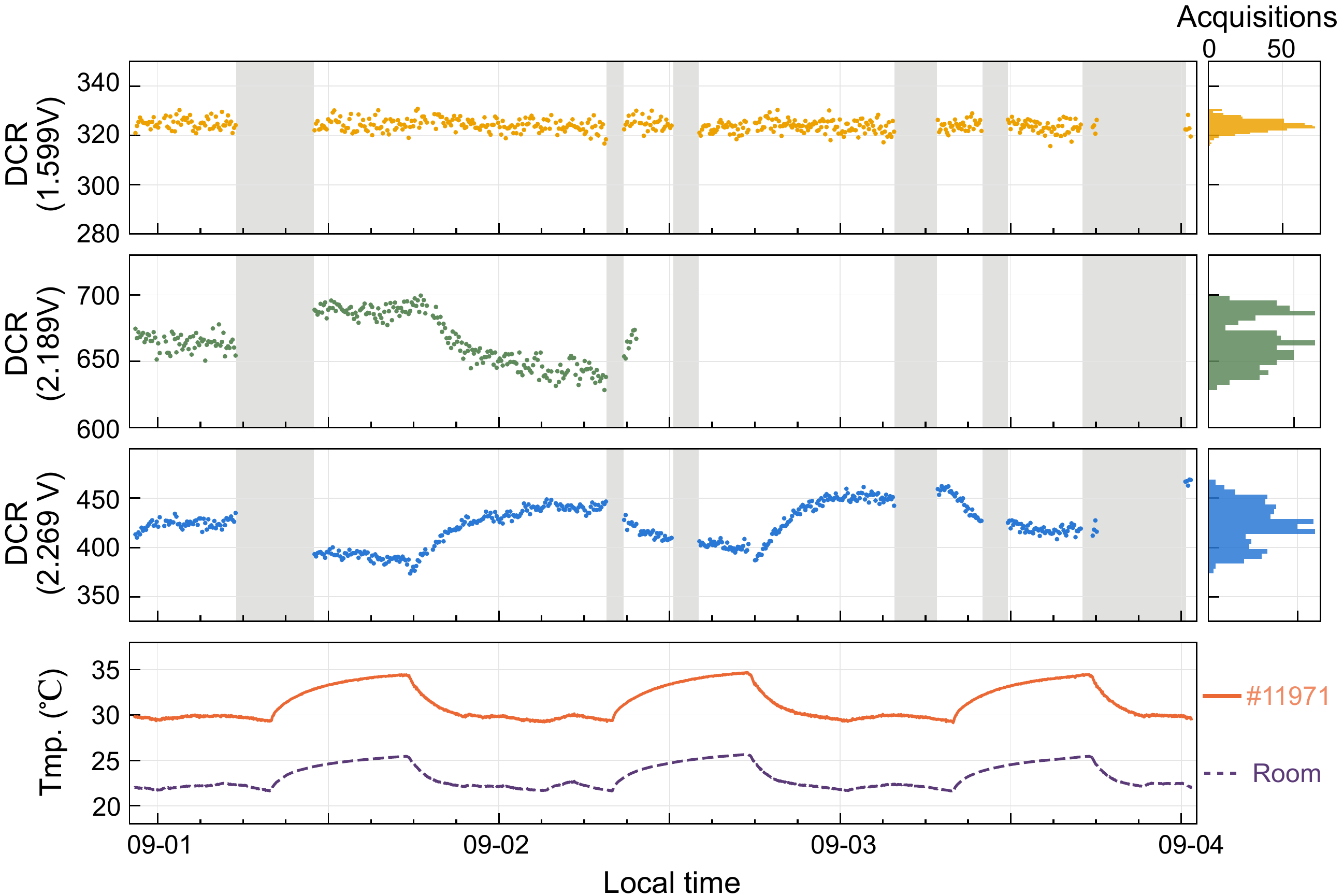}}
\caption{Count rates of \#11971 at three TDC thresholds with a 6\,dB attenuator, together with the module-housing and room temperatures. Only acquisitions in the intrinsic high-DCR state are analyzed; grey bands indicate excluded intrinsic low-state intervals. Right: marginal rate histograms.}
\label{fig:TmpArt}
\end{figure}

In a three-day experiment, the TDC threshold was stepped cyclically while the detector remained sealed in darkness, and the module-housing and room temperatures were recorded continuously. Fig.~\!\ref{fig:TmpArt} shows the rates measured at the normal operating threshold of 1.599\,V and at two thresholds near the attenuated pulse top, 2.189\,V and 2.269\,V. Only acquisitions in the intrinsic high-DCR state, identified from the operating-threshold record, were retained, while intrinsic low-state intervals are shaded in grey and excluded. By restricting the analysis to a single intrinsic DCR state, any apparent switching observed at the pulse-top thresholds can be attributed to the temperature-related threshold-crossing effect rather than to intrinsic RTS, demonstrating that this mechanism alone can generate RTS-like rate variations.

As observed in Fig.~\!\ref{fig:TmpArt}, the count rate at the normal operating threshold remains near 324\,Hz, varying by less than 0.4\% throughout the temperature cycles, and exhibits a single-peaked distribution. In contrast, the rates measured near the pulse top show pronounced temperature dependence. Specifically, at 2.189\,V, where the two local maxima of a single output pulse can cross the threshold separately, the recorded rate is approximately twice the plateau rate and decreases as the module cools. At 2.269\,V, the response is reversed, with the rate increasing upon cooling. Across three temperature steps in two independent runs, the 2.269\,V rate increased by 8.2\% and 8.0\% for cooling steps of 3.4 and 3.6\,K, respectively, and decreased by 7.5\% for a warming step of 3.0\,K. These observations show that the apparent rate changes closely track the temperature and reverse with the direction of the temperature change. As a result, the marginal rate distribution can become bimodal even when the detector remains in a single intrinsic DCR state, demonstrating that temperature-related threshold crossings can produce RTS-like behavior.

\begin{figure}[htb]
\centerline{\includegraphics[width=\linewidth]{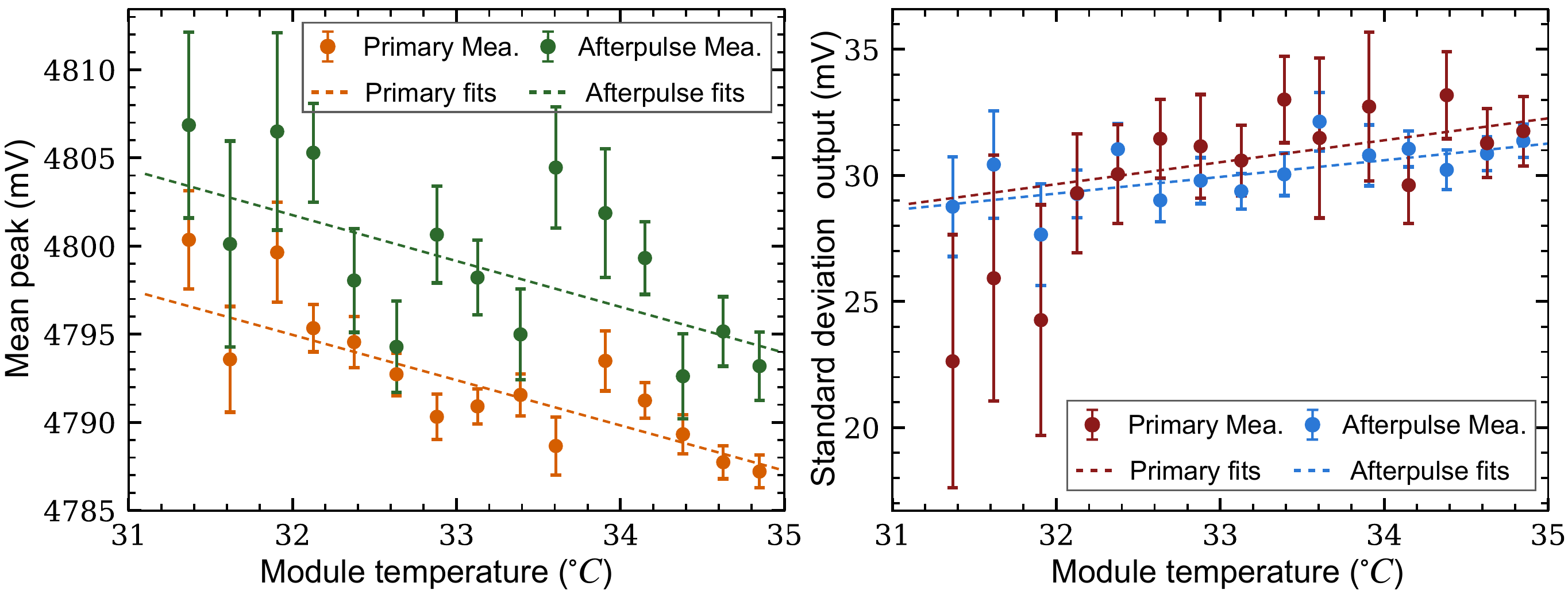}}
\caption{Peak amplitudes of primary pulses and afterpulses of \#11971 versus module temperature during warming. (Left) Mean peak amplitude. (Right) Standard deviation. Dashed lines show weighted least-squares fits.}
\label{fig:AmpTmp}
\end{figure}

The origin of this artifact was further examined using oscilloscope measurements. In total, 36,441 primary--afterpulse pulse pairs were recorded. During a warming interval from 31.2 to 35.0\,$^\circ$C, 8,462 pulses were analyzed. As shown in Fig.~\!\ref{fig:AmpTmp}, the mean peak amplitude of the primary pulses changed by $-2.57\pm0.35$\,mV/K. After correcting for the accidental-pair fraction of 0.47, the afterpulse amplitude showed a similar temperature coefficient of $-2.60\pm0.71$\,mV/K. In comparison, the pulse-to-pulse standard deviation remained within 28--33\,mV and changed only weakly with temperature, by $+0.66\pm0.24$\,mV/K for primary pulses and $+0.87\pm0.52$\,mV/K for afterpulses. These measurements show that temperature mainly shifts the pulse-height distribution by a few millivolts per kelvin without substantially changing its width. Such a shift has little effect at the normal operating threshold, which lies well below the pulse top, but can noticeably change the number of pulses crossing a threshold placed near the amplitude flank.

\begin{figure}[htb]
\centerline{\includegraphics[width=\linewidth]{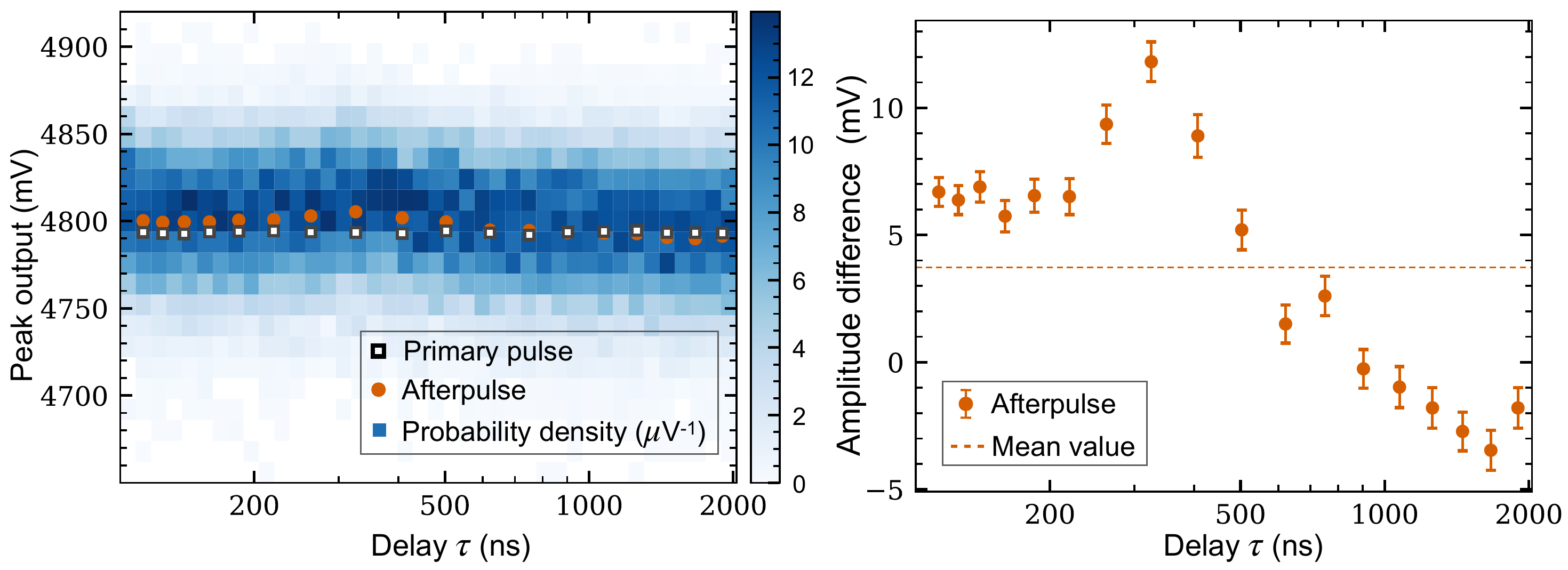}}
\caption{Peak amplitudes of primary and afterpulse TDC outputs versus delay. (Left) Afterpulse amplitude distribution with the corresponding primary- and afterpulse-amplitude means. (Right) Afterpulse--primary amplitude difference.}
\label{fig:AmpDis}
\end{figure}

Although the SPAD module provides a digital transistor--transistor logic (TTL) output, the pulse amplitude is not perfectly constant and differs between a primary pulse and its associated afterpulse in a delay-dependent manner. As shown in Fig.~\!\ref{fig:AmpDis}, the afterpulse is, on average, approximately 6.7\,mV higher than its associated primary at short delays, and the difference gradually increases to a maximum of 11.8\,mV at a delay of 326\,ns before gradually decreasing. These results indicate that the amplitudes of a primary pulse and its afterpulse are coupled, but not identical. When the discriminator threshold is placed near the pulse top, the residual amplitude difference causes the two pulses to have different probabilities of crossing the threshold. Their detection losses are therefore neither independent nor perfectly correlated, leading to an intermediate pair-survival response.

We next apply the self-referenced diagnostic to the temperature-related threshold-crossing artifact, with the results summarized in Table~\ref{tab:TmpArt}. At the normal operating threshold of 1.599\,V, the two temperature phases are nearly indistinguishable in count rate ($p=0.997$), indicating that the temperature variation does not produce a meaningful DCR transition under normal operation. In contrast, at 2.269\,V and 2.309\,V, the apparent rate changes reach approximately $9.9\%$ and $31.0\%$, respectively.

\begin{table}[htb]
\centering
\scriptsize
\caption{Diagnosis of the temperature-related threshold-crossing artifact of \#11971, compared with its intrinsic RTS at the operating threshold.}
\label{tab:TmpArt}
\setlength{\tabcolsep}{4pt}
	\begin{tabular}{cccc}
		\hline
		Case & $p$ & $R$ & $\rho_s$\\
		\hline
		Artifact, 1.599\,V & 0.997 & $1.026\pm0.024$ & ---\\
		Artifact, 2.269\,V & 0.901 & $0.959\pm0.013$ & $0.59\pm0.13$\\
		Artifact, 2.309\,V & 0.690 & $0.842\pm0.036$ & $0.49\pm0.12$\\
		Intrinsic RTS, 1.599\,V & 0.554 & $0.9996\pm0.0027$ & $0.999\pm0.006$\\
		\hline
	\end{tabular}
\end{table}

For both threshold-induced artifacts, the measured afterpulse-yield ratio lies between the two limiting cases of independent and fully correlated loss. In detail, at 2.269\,V, $R=0.959\pm0.013$ lies between $p=0.901$ and unity, giving $\rho_s=0.59\pm0.13$. Similarly, at 2.309\,V, $R=0.842\pm0.036$ lies between $p=0.690$ and unity, giving $\rho_s=0.49\pm0.12$. Thus, the threshold-crossing loss is neither independent, for which $R=p$ and $\rho_s=0$, nor perfectly correlated, for which $R=1$ and $\rho_s=1$. Instead, the intermediate values indicate partial coupling between the detection of a primary pulse and that of its associated afterpulse, consistent with their correlated but nonidentical amplitudes.

This behavior is clearly different from the intrinsic RTS of the same detector. At the normal operating threshold, the intrinsic transition of SPCM-AQR-14 \#11971 gives $R=0.9996\pm0.0027$ despite a much larger rate ratio of $p=0.554$, corresponding to $\rho_s=0.999\pm0.006$. Therefore, the intrinsic RTS preserves the afterpulse yield across the two DCR states, whereas the temperature-related threshold-crossing artifact produces a reduced $R$ and an intermediate $\rho_s$. The combination of $R$ and $\rho_s$ therefore separates this artifact from intrinsic RTS even though both can produce apparent two-level changes in the recorded count rate.

%% file: EnvArt.tex
\section{Environment-Induced Artifacts}
\label{sec:EnvArt}
This section examines RTS-like DCR step changes caused by variations in the measurement environment, including ambient-light leakage and cable-related loss. These artifacts can reproduce several features of intrinsic RTS and therefore cannot always be identified from the afterpulse-yield ratio $R$ alone. Instead, they can be distinguished by combining the self-referenced metrics with characteristic features observed directly in the DCR record.

\subsection{RTS-Like Artifacts Caused by Light Leakage}
\label{subsec:LightLeak}

\begin{figure}[htb]
	\centerline{\includegraphics[width=\linewidth]{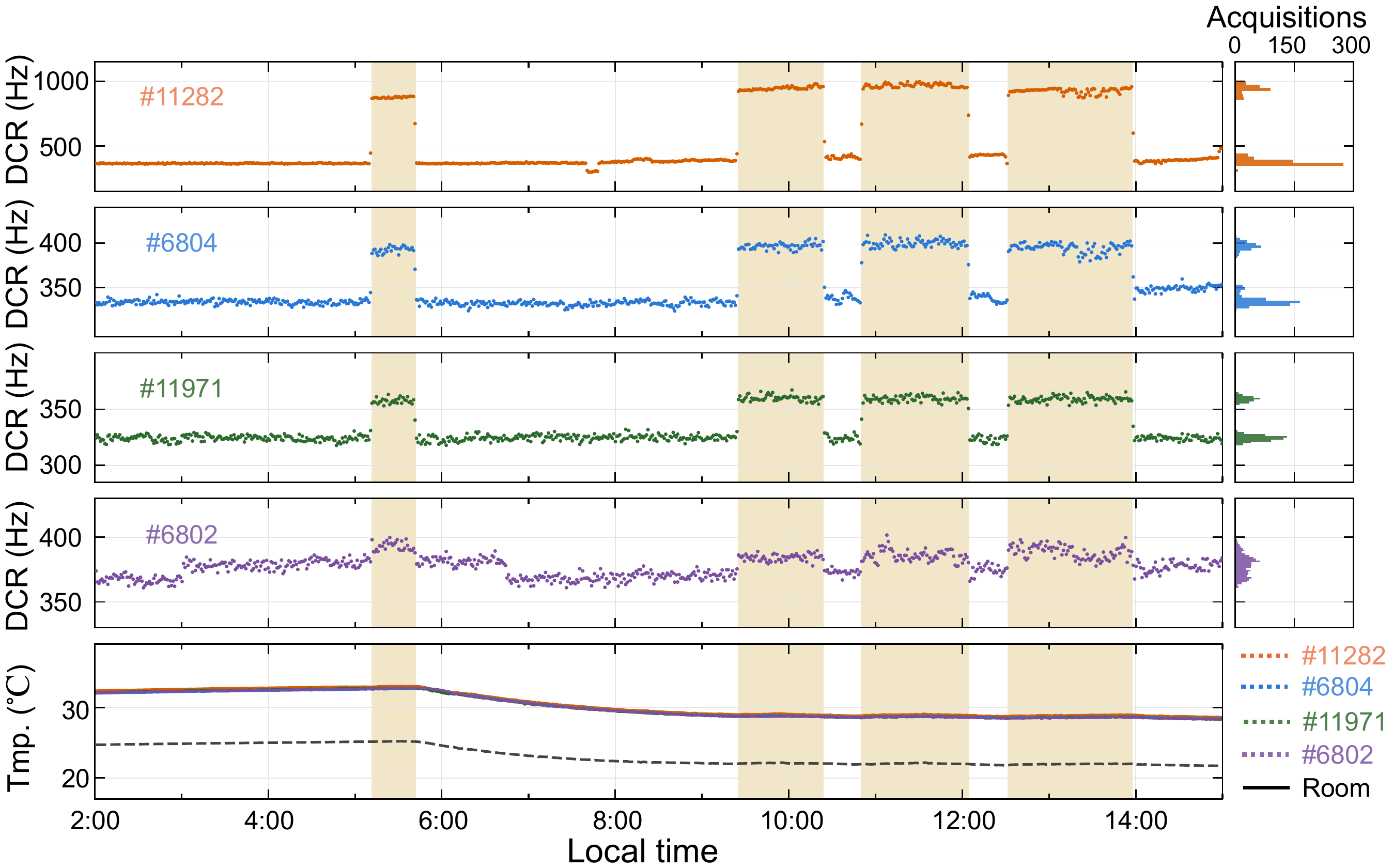}}
	\caption{Light-leak episode on 08-26 from 02:00 to 15:00 CDT. The top four panels show the DCRs of the four SPAD modules and their marginal histograms under light-leakage conditions, while the bottom panel shows the module-housing and room temperatures. Yellow shaded intervals indicate periods when the laboratory ceiling lights were on.}
	\label{fig:LightLeak}
\end{figure}

Light leakage can produce abrupt two-level changes in the recorded DCR that closely resemble intrinsic RTS. During the initial measurements, the SPAD housings were not fully light-tight, allowing laboratory illumination to reach the detectors. Fig.~\!\ref{fig:LightLeak} shows a 13\,h record from 02:00 to 15:00 on 08-26, containing four illuminated periods and eight light-switching edges. All four modules change rate simultaneously, with mean light-induced steps of approximately 543\,Hz for \#11282, 61\,Hz for \#6804, 35\,Hz for \#11971, and 13\,Hz for \#6802. The different step amplitudes reflect different degrees of light leakage into the individual modules. When viewed on a single channel, these abrupt two-level changes can closely resemble intrinsic RTS.

The light-induced count-rate transitions are also very sharp. For \#11282, fitting the eight switching edges with the same two-level Poisson model used for the intrinsic RTS yields no resolvable finite transition ramp, with a pooled 95\% upper limit of $\tau<0.42$\,ms. This shows that transition sharpness alone cannot distinguish a light-leakage artifact from intrinsic RTS.

A useful distinction emerges from the DCR levels themselves. Unlike the reproducible levels of intrinsic RTS, the apparent high- and low-rate states produced by light leakage are not fixed. For \#11282, the dark level adjacent to the switching edges varies from 359.5 to 399.9\,Hz, while the illuminated level varies from 867.8 to 956.2\,Hz. The corresponding light-induced DCR step also varies, from approximately 508 to 566\,Hz. These changes can be separated into two optical contributions: a slowly varying daylight background and variations in the output of the ceiling lamps, as discussed below.

Daylight shifts both the dark and illuminated DCR levels and therefore appears mainly as a common baseline drift. Because the lamp contribution is obtained from the rate difference immediately across each switching edge, this slowly varying daylight component largely cancels. For \#11282, during the 09:24--10:24 interval, the illuminated level increases by 24.3\,Hz while the adjacent dark level increases by 21.7\,Hz, whereas the switch-on and switch-off steps differ by only 2.6\,Hz. Similarly, during 10:50--12:04, the illuminated and dark levels change by 8.5 and 8.8\,Hz, respectively, while the two lamp steps differ by only 0.3\,Hz. The common drift of the two levels can therefore be attributed primarily to daylight, whereas their difference isolates the ceiling-lamp contribution.

\begin{figure}[htb]
	\centerline{\includegraphics[width=\linewidth]{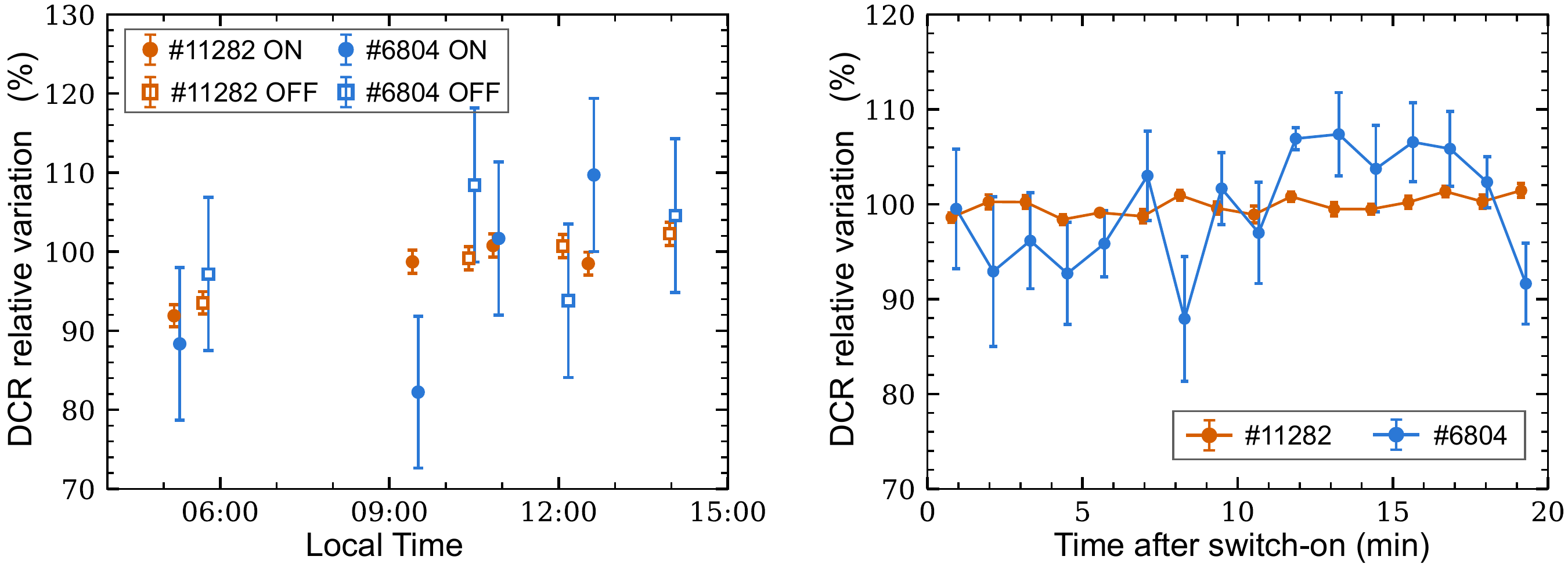}}
	\caption{Variation of the light-induced DCR contribution for \#11282 and \#6804. (Left) Light-induced step at successive switch-on and switch-off edges, normalized to the daytime mean of each module. (Right) Evolution of the light contribution after the 05:11 switch-on, normalized to the mean over the corresponding illuminated period.}
	\label{fig:LampVar}
\end{figure}

An important observation is that the lamp contribution is not perfectly constant, as the lamp output varies with the thermal condition of the fixtures. As shown in Fig.~\!\ref{fig:LampVar}, the six daytime steps of \#11282 measured between 09:24 and 13:58 average $553.1\pm3.3$\,Hz, whereas the first switch-on/off pair at 05:11 and 05:41 averages only $512.8\pm5.6$\,Hz, or $7.3\%$ lower. The same relative reduction is observed independently on \#6804, whose first-period step is $92.7\pm7.8\%$ of its daytime value, compared with $92.7\pm1.1\%$ for \#11282. Because the two modules have different degrees of light leakage, their agreement indicates that the reduction originates from the common light source rather than from a change in either detector.

The smaller lamp-induced DCR step observed during the first illuminated period, from 05:11 to 05:41, is associated with the thermal state of the lamp fixtures, as shown in Fig.~\!\ref{fig:LampVar}. At 05:11, the room temperature was approximately 25.1\,$^\circ$C, compared with 21.8--22.1\,$^\circ$C during the later switching events. This difference cannot be explained simply by a cold start: although the lamps had been off for 3.7\,h before the 09:24 switch-on, the resulting step recovered to the normal daytime value of 553.1\,Hz. The reduced first-period step is therefore more consistent with the warmer ambient condition of the fixtures than with the time since the lamps were switched on.

A smaller warm-up effect is also visible after the 05:11 transition. For \#11282, the lamp contribution increases from $500.1\pm1.7$\,Hz during 0.5--2\,min after switch-on to $508.7\pm1.2$\,Hz during 10--25\,min. After accounting for the concurrent baseline drift, the lamp itself increases by only about $1\%$, much smaller than the $7.3\%$ difference between the first illuminated period and the later daytime measurements.

These variations in the DCR provide a direct way to distinguish light-leakage artifacts from intrinsic RTS. In intrinsic RTS, the two DCR levels are reproducible over repeated transitions. In contrast, light leakage produces apparent states whose absolute levels shift with the daylight background and whose separation changes with the thermal condition of the external light source. The artifact is further identified by the simultaneous transitions across multiple detector channels, whereas intrinsic RTS is detector-local. Thus, even when a single-channel record appears bimodal and its transitions are abrupt, the cross-channel coincidence and the lack of reproducible DCR levels reveal an external optical origin.

The afterpulse measurement alone cannot make this distinction. For \#11282, the light-induced states give $p=0.419$, $R=1.015\pm0.012$, and $\rho_s=1.03\pm0.02$. The near-unity $R$ strongly rejects both independent counting loss and uncorrelated downstream count addition. Similar near-unity values are obtained for \#6804 and \#11971. This behavior is expected because leaked photons generate genuine SPAD avalanches, and these avalanches retain their normal probability of producing afterpulses. Light leakage can therefore produce the same $R\simeq1$ and $\rho_s\simeq1$ signature as intrinsic RTS. In this case, the distinction is provided by the DCR record itself: simultaneous switching across channels, externally determined dwell intervals, and rate levels that vary with daylight and lamp thermal conditions identify the apparent RTS as light leakage rather than intrinsic detector switching.

\subsection{Cable-Induced Counting Artifacts}
\label{sec:CableArtifact}

Mechanical disturbances in the signal path can also produce RTS-like changes in the recorded DCR. A loose connector, intermittent contact, or stressed cable can generate additional electrical triggers or temporarily interrupt the signal without changing the underlying SPAD avalanche process. Such artifacts may therefore appear as abrupt count-rate changes and be mistaken for intrinsic detector switching if only the averaged DCR is considered.

\begin{figure}[htb]
    \centering
    \includegraphics[width=\linewidth]{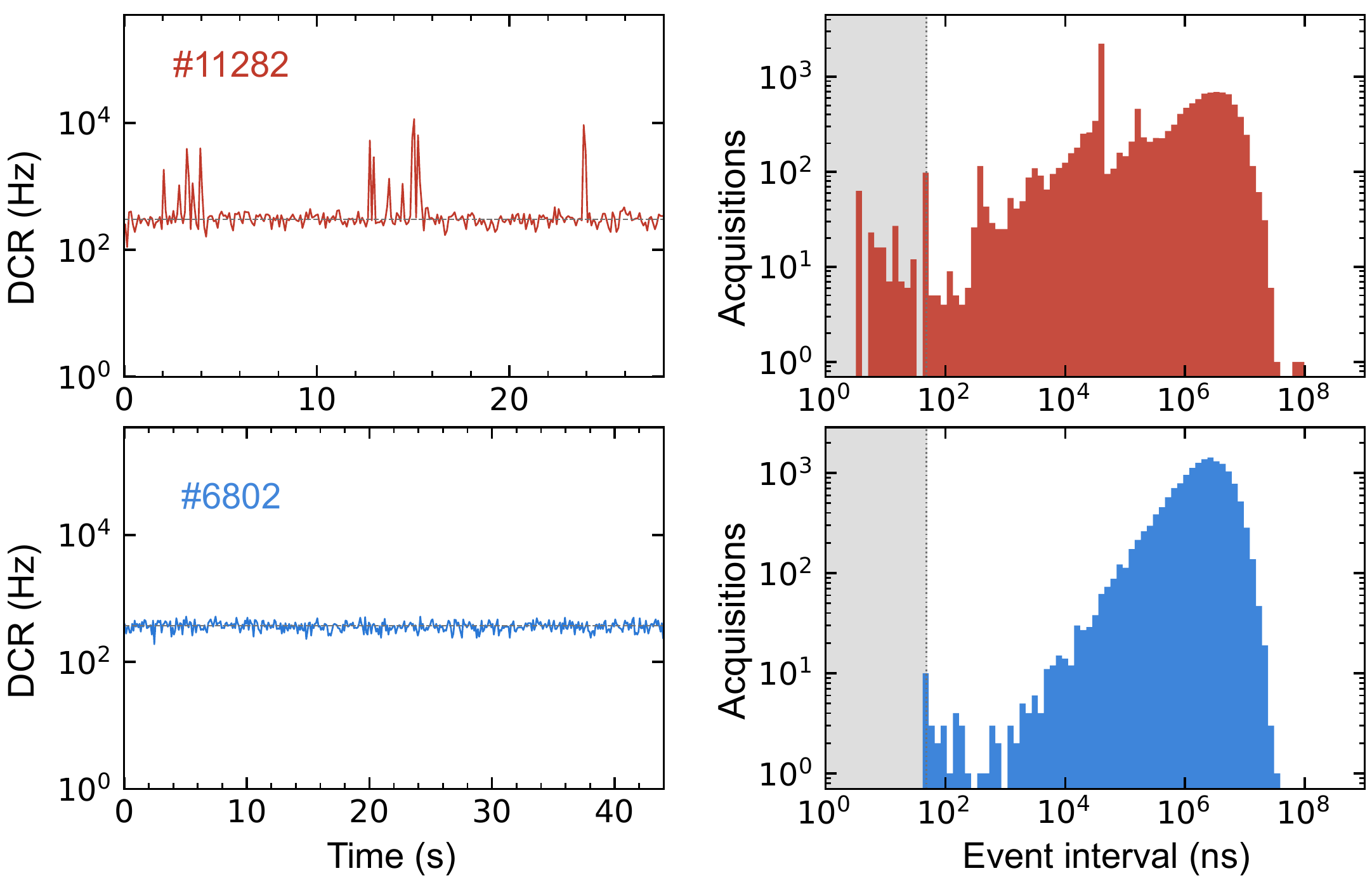}
    \caption{Comparison of a cable-contact artifact and normal detector operation. (Left) Count-rate traces. (Right) Inter-event-interval distributions, where the gray shaded region indicates the SPAD dead-time interval.}
    \label{fig:CableTwist}
\end{figure}

To reproduce this effect, the output connection of module \#11282 was deliberately twisted during continuous acquisition to induce intermittent contact glitches. Module \#6802 under normal operation was recorded as a reference. The resulting count-rate traces and inter-event-interval distributions are shown in Fig.~\!\ref{fig:CableTwist}. As illustrated in the figure, cable twisting does not produce a stable additional rate state. Instead, the DCR fluctuates around a baseline of approximately 300\,Hz, with irregular bursts reaching several kilohertz and occasional signal dropouts. In contrast, the normally operating detector remains near its approximately 370\,Hz baseline without comparable excursions. This behavior differs from intrinsic RTS, which is characterized by stochastic switching between reproducible and relatively stable DCR levels.

The distinction becomes even clearer in the inter-event timing. Under normal operation, the SPCM-AQR-14 \#11282 module exhibits a 48\,ns dead-time floor, with no events observed below this limit. During cable twisting, however, the minimum interval decreases to 4\,ns, and approximately 1.90\% of the intervals fall below 48\,ns. Such events cannot originate from genuine SPAD avalanches because the detector cannot produce another physical avalanche during its recovery time. They are therefore attributed to electrical multi-triggering or contact-induced ringing in the readout path.

The afterpulse statistics provide an additional discriminator. For the clean intrinsic RTS of \#11282, we obtain $R=1.008$ with $p=0.810$, corresponding to $\rho_s=1.04$, consistent with the intrinsic-switching prediction. During cable twisting, the same analysis gives $R=1.283$ with $p=0.822$ and $\rho_s=2.59$. The elevated $R$ arises from contact-induced multi-triggering that adds correlated counts within the afterpulse analysis window. The resulting $\rho_s$ lies well outside the pure-loss range $0\leq\rho_s\leq1$, providing further evidence that the disturbance is not intrinsic RTS.

Thus, cable-induced counting artifacts can be distinguished from intrinsic RTS through both their rate and timing characteristics. Cable faults produce irregular bursts, signal dropouts, and events within the SPAD-forbidden sub-dead-time region, whereas intrinsic RTS preserves reproducible rate levels and the detector's physical dead-time constraint. The combination of abnormal $R$ and $\rho_s$ values with sub-dead-time events therefore provides a clear signature of a non-intrinsic cable artifact that can be readily identified using the proposed self-referenced diagnostic.

%% file: ReadArt.tex
\section{Readout-Dependent Artifacts}
\label{sec:ReadArt}

\subsection{Injection-Induced Counting Addition}
\label{sec:InjCount}
\subsubsection{Artificial Injection}
Artificial signals injected into the electronic readout path can alter the recorded output of a quantum measurement system without modifying the underlying quantum process. Such electronic signal-injection attacks have been demonstrated in practical quantum hardware, where externally coupled signals can manipulate the analog readout and affect the recorded result~\cite{pantoja2024electromagnetic, smith2021out}. In an SPAD quantum communication receiver, an injected pulse stream that is switched on and off could similarly add counts to the detector output and produce an apparent two-level DCR change resembling RTS.

A more challenging case arises when the injected counts are deliberately designed to resemble the target detector. An attacker with prior access to the SPAD could, in principle, measure its DCR and afterpulse statistics, including the afterpulse probability and delay distribution, and then use these statistics to construct artificial primary--afterpulse pairs. If the injected population reproduces the same afterpulse yield as the detector, the resulting rate change can satisfy $R\simeq1$, making the afterpulse-yield ratio alone unable to distinguish the artificial counts from intrinsic RTS.

To examine this limiting case, we implemented a controlled electrical injection that reproduces the measured afterpulse statistics of a real detector. The purpose of this experiment is not to reproduce a particular reported attack, but to test the proposed diagnostic under a deliberately difficult condition in which the artificial counts are constructed to resemble the detector as closely as possible in their DCR and primary--afterpulse statistics.

The experiment was performed using module \#6802. A Keysight 33622A arbitrary waveform generator (AWG) generated the artificial pulse stream, which was combined with the SPCM output through a Wilkinson power divider before being recorded by the TDC. The AWG operated at 40\,MS/s, corresponding to 25\,ns per sample, and generated primary pulses at approximately 90\,Hz. Each injected primary was followed, with a probability of 0.228\%, by a secondary pulse whose delay was randomly sampled from the measured afterpulse-delay distribution of \#6802 over 50\,ns to 20,000\,ns. The follower probability was chosen to match the measured afterpulse yield of the detector, and a new random sequence was loaded every 45\,s so that the injected delays sampled the measured distribution throughout the acquisition.

\begin{figure}[htp]
    \centering
    \includegraphics[width=\linewidth]{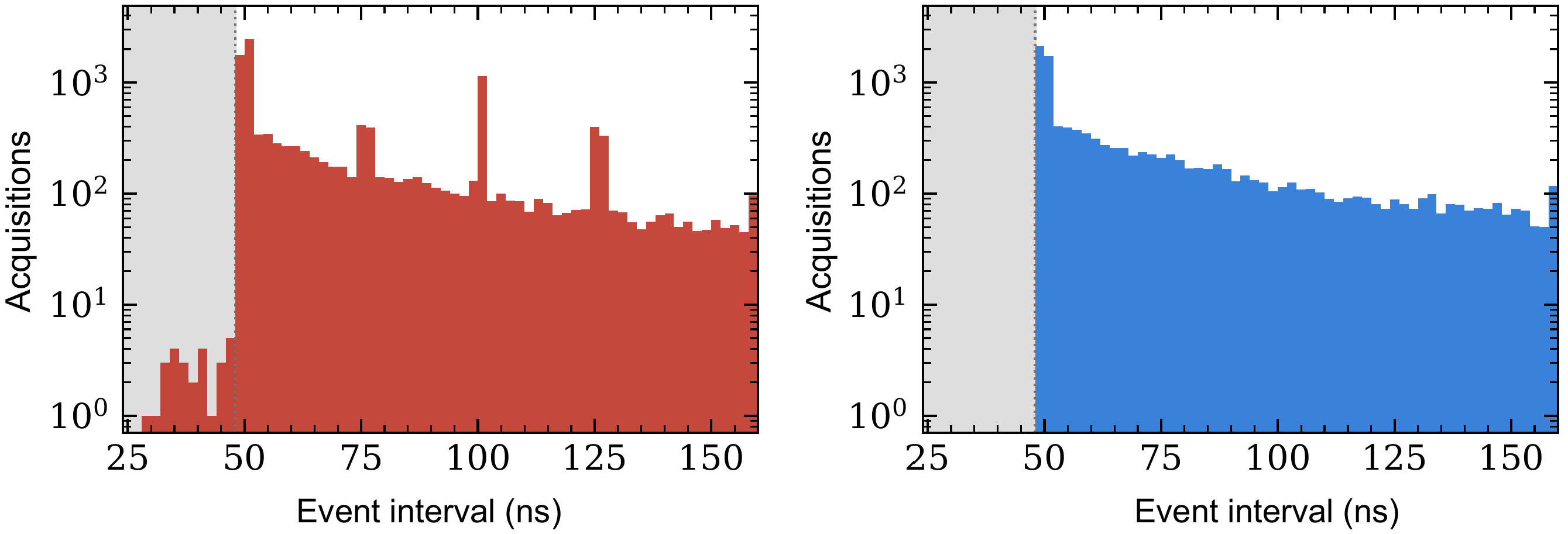}
    \caption{Dead-time signature of the artificial pulse injection (left) and normal detector operation (right). }
    \label{fig:InjDeadTime}
\end{figure}

As intended, the artificial population reproduces the detector afterpulse yield. Over the full analysis window, the measured ratio is $R=1.002\pm0.020$, with $\rho_s\simeq1$. The injected counts therefore occupy the same region as intrinsic RTS in the $R$--$\rho_s$ test. Their artificial origin, however, remains visible in the inter-event timing. As shown in Fig.~\!\ref{fig:InjDeadTime}, the real detector has a hard minimum inter-event interval of 48\,ns (gray shaded region in the figure) on \#6802, with no events below this limit in more than $3.0\times10^{7}$ measured intervals. In contrast, the injection produces 40 events in the 28--48\,ns region, with a minimum interval of 28\,ns.

These results show that reproducing the DCR and afterpulse statistics is not sufficient to reproduce the full temporal behavior of a real SPAD. Because the injected electrical pulses bypass the detector avalanche and recovery process, they can occur within the dead-time region that is inaccessible to genuine detector events. Consequently, a deliberately matched injection can produce $R\simeq1$ and $\rho_s\simeq1$, making the self-referenced yield test alone inconclusive, while the dead-time correlation still reveals its non-intrinsic origin. Mimicking intrinsic RTS would therefore require reproducing not only the count rate and afterpulse statistics, but also the physical recovery constraint of the detector.

\subsubsection{Interference-Induced Random Counting Addition}
\label{sec:RanAdd}
Another readout artifact is the random addition of spurious counts unrelated to the SPAD avalanche process. Electrical noise or interference in the readout chain can cross the discrimination threshold and be registered as detection events. This behavior reflects the practical trade-off among discrimination threshold, detection efficiency, and electronic noise in SPAD systems~\cite{wang2012detection,zhang2015advances,liu2022double}. If the interference level or effective threshold shifts between two operating conditions, the resulting change in the false-count rate can produce an RTS-like DCR transition.

A direct example was reported in \cite{restelli2014ultra} for a 1.25\,GHz InGaAs/InP SPAD system. At a sufficiently low discrimination threshold, false detections occurred with probabilities of order $10^{-6}$ per gate even when the SPAD was biased below breakdown. Because avalanches cannot occur under this condition, these events were attributed to the comparator triggering on electrical-noise fluctuations. Similar threshold dependence has been reported in other InGaAs/InP SPAD systems: lowering the discriminator threshold improves the acceptance of small avalanche signals but also increases the contribution from electrical noise~\cite{wang2012detection}. More generally, comparator thresholds must be kept above the residual electronic background to suppress false detections~\cite{zhang2015advances,liu2022double}.

For random counts added downstream of the SPAD, the added events increase the singles rate without contributing the corresponding primary--afterpulse pairs. If the added counts are uncorrelated with the detector events, the expected afterpulse-yield ratio is $R=1/p$. For example, a $10\%$ increase in the registered rate due entirely to uncorrelated added counts gives $p=1/1.1\approx0.91$ and $R=1/p\approx1.10$. The corresponding normalized statistic is $\rho_s\approx2.1$, well outside the pure-loss range $0\leq\rho_s\leq1$. This behavior is therefore clearly separated from intrinsic RTS, for which the avalanche population changes while the afterpulse yield remains approximately unchanged, giving $R\simeq1$.

The inter-event timing also provides an independent diagnostic. Because downstream electronic counts do not originate from SPAD avalanches, they are not inherently constrained by the physical SPAD recovery time and may appear below the detector dead-time floor if the readout electronics can resolve them. The combination of $R>1$, $\rho_s>1$, and sub-dead-time events therefore provides strong evidence for interference-induced counting addition. If the interference is temporally correlated or produces bursts rather than independent counts, the ideal relation $R=1/p$ no longer applies exactly; nevertheless, the resulting deviations in $R$ and the violation of the detector dead-time structure can still distinguish the artifact from intrinsic RTS.

\subsection{Electronics-Induced Counting Loss}
\label{sec:EleCount}

In practical SPAD systems, genuine avalanches can be selectively lost in the readout electronics. Changes in the discriminator or signal-processing conditions can therefore alter the registered count rate without changing the underlying avalanche population. If the readout switches between two stable conditions, the resulting DCR can resemble intrinsic RTS.

One example was reported in \cite{wayne2021low}, where the detection efficiency of a high-rate Si SPAD system decreased at count rates above approximately $2.5\times10^{7}\,\mathrm{s}^{-1}$. At $10^{8}\,\mathrm{s}^{-1}$, the measured efficiency decreased to $0.45\pm0.01$, compared with a low-rate value of approximately 0.543. This reduction was attributed to AC coupling between the output amplifier and the discrimination comparator. As the count rate increases, the resulting duty-cycle-dependent baseline shift changes the effective discrimination threshold, causing some genuine avalanche pulses to be missed. If these two efficiencies are taken as representative of two hypothetical readout states, the corresponding rate ratio is $p = 0.45/0.543 \approx 0.83$.

The corresponding values of $R$ and $\rho_s$ depend on whether a primary pulse and its associated afterpulse are lost independently or together. If the effective threshold remains approximately fixed within each state and the pulse amplitudes fluctuate independently, the two losses are also approximately independent. In this limit, $R\simeq p\simeq0.83$ and $\rho_s\simeq0$. The AC-coupled baseline, however, can introduce correlation between the two survival decisions. A primary pulse and an afterpulse separated by a sufficiently short interval may experience similar local baseline and threshold conditions, making the detection of the two pulses positively correlated. Under positive common-mode coupling, the response moves toward $p<R<1$ and $0<\rho_s<1$. Thus, this mechanism is neither inherently independent nor fully coupled; the degree of coupling depends on the temporal behavior of the AC-coupled baseline and on the pulse-amplitude statistics.

A more complex example was reported in \cite{su2026security}. In a 1\,GHz SPAD receiver equipped with a pulse-width discriminator, avalanche responses from adjacent events can merge into a wider pulse. If the pulse width exceeds the accepted discrimination window, the avalanche is rejected as a valid click even though it has already occurred and the detector still enters its normal dead time. Under the reported condition, the registered count rate decreased from approximately 150\,Hz to 32\,Hz, giving $p = 32/150 \approx 0.21$.

This process cannot be described as simple independent thinning. A rejected wide avalanche still changes the subsequent detector state, and the reported residual counts include afterpulses appearing after recovery as well as clicks associated with the avalanche tail. Whether a later event is recorded therefore depends on both the preceding avalanche history and its temporal separation from that avalanche. The pair-survival probability is consequently related to the delay between the primary pulse and its associated afterpulse, i.e., $\Delta t$, rather than being described by a constant survival probability. As a result, the process falls outside the pure-loss model discussed in Sec.~\ref{sec:NorPair}. In this case, $\rho_s$ serves only as a normalized diagnostic statistic, and the published data are insufficient to determine the corresponding $R$.

These two examples illustrate how the proposed diagnostic responds to readout-induced RTS artifacts. For selective counting loss with independent or partially correlated primary--afterpulse survival, the afterpulse yield changes with the registered rate, giving $R<1$ and distinguishing the artifact from the intrinsic RTS observed here, for which $R\simeq1$. This distinction becomes weaker as the two survival decisions become more strongly coupled, since nearly common loss can also produce $R\simeq1$. For history- and delay-dependent processes such as pulse-width discrimination, $R$ and $\rho_s$ alone may not be sufficient because the observed event stream is no longer a simple thinned version of the underlying avalanche stream. In such cases, the self-referenced quantities should be interpreted together with the time-tagged DCR behavior. Readout-induced artifacts may show rate levels or temporal structures that depend on the readout process, whereas intrinsic RTS is characterized by reproducible detector-specific rate states and stochastic switching between them.

%% file: Conclusion.tex
\section{Conclusion}
\label{sec:Con}

This work demonstrates that an apparent DCR transition cannot be reliably identified from the count-rate trace alone. The intrinsic RTS observed in the tested SPAD quantum receivers is reproducible, detector-specific rate levels, approximately memoryless dwell times, and abrupt switching with no resolvable transition ramp. Although the absolute afterpulse yield varies across devices and with time and temperature, the ratio of the yield between the resolved DCR states remains stable. This stability provides the self-reference underlying the proposed diagnostic approach.

The controlled-artifact measurements further show that different mechanisms produce distinct correlation and timing signatures. For temperature-induced threshold crossings, the measured responses yield $R=0.959\pm0.013$ with $\rho_s=0.59\pm0.13$ at 2.269\,V and $R=0.842\pm0.036$ with $\rho_s=0.49\pm0.12$ at 2.309\,V. These values differ clearly from those of the intrinsic RTS in \#11971, for which $R=0.9996\pm0.0027$ and $\rho_s=0.999\pm0.006$. Cable-contact disturbances provide another distinguishable signature, producing anomalous short-delay events below the SPAD dead-time floor.

Importantly, however, $R\simeq1$ is not by itself unique to intrinsic RTS. Light leakage introduces genuine additional avalanches and can therefore preserve the afterpulse yield. Similarly, an artificial electrical injection deliberately matched to the measured afterpulse statistics gives $R=1.002\pm0.020$ with $\rho_s\simeq1$. Despite this apparent agreement with intrinsic RTS, the injected signal produces events below the SPAD dead-time floor, revealing its non-intrinsic origin. By contrast, the RTS-like behavior induced by light leakage varies with temperature and with the thermal state of the light source, providing an additional means of distinguishing it from the reproducible intrinsic RTS. Taken together, these results show that $R$ and the associated $\rho_s$ are most informative when interpreted alongside the temporal structure and reproducibility of the DCR record. The resulting framework provides a practical approach for distinguishing detector-local RTS from optical, electrical, and readout-induced artifacts using primarily time-tagged dark-count measurements.